\documentclass{aa}

\usepackage{graphicx}
\usepackage{txfonts}
\usepackage{color}
\definecolor{darkblue}{rgb}{0.00,0.00,0.40}
\usepackage[colorlinks=true,citecolor=darkblue]{hyperref}

\usepackage{xspace}
\newcommand{\kms}{\,km\,s$^{-1}$\xspace}
\newcommand{\hiireg}{\textsc{H\,ii}~region\xspace}
\newcommand{\hiiregs}{\textsc{H\,ii}~regions\xspace}
\newcommand{\ha}{H$\alpha$\xspace}
\newcommand{\hb}{H$\beta$\xspace}
\newcommand{\pazeta}{Pa$\zeta$\xspace}

\newcommand{\hei}{\ion{He}{i}\xspace}

\newcommand{\Te}{\ensuremath T_\mathrm{e}\xspace}
\newcommand{\Ne}{\ensuremath N_\mathrm{e}\xspace}
\newcommand{\chb}{\ensuremath c_\mathrm{H\beta}\xspace}

\newcommand{\oi}{[\ion{O}{i}]\xspace}
\newcommand{\oii}{[\ion{O}{ii}]\xspace}
\newcommand{\nii}{[\ion{N}{ii}]\xspace}
\newcommand{\sii}{[\ion{S}{ii}]\xspace}
\newcommand{\oiii}{[\ion{O}{iii}]\xspace}
\newcommand{\siii}{[\ion{S}{iii}]\xspace}
\newcommand{\cliii}{[\ion{Cl}{iii}]\xspace}

\newcommand{\lr}{{\tt LR|1.25}\xspace}
\newcommand{\hr}{{\tt HR|0.85}\xspace}

\begin{document}
\title{A MUSE map of the central Orion Nebula (M\,42)
       \thanks{Data products are available at \url{http://muse-vlt.eu/science}.}
       \fnmsep
       \thanks{Based on observations made with ESO telescopes at the La Silla
               Paranal Observatory under program ID 60.A-9100(A).}
}
\author{Peter M.\ Weilbacher\inst{1}
        \and
        Ana Monreal-Ibero\inst{2}
        \and
        Wolfram Kollatschny\inst{3}
        \and
        Adam Ginsburg\inst{4}    
        \and
        Anna F.\ McLeod\inst{4}  
        \and
        Sebastian Kamann\inst{3} 
        \and
        Christer Sandin\inst{1}  
        \and
        Ralf Palsa\inst{4}       
        \and
        Lutz Wisotzki\inst{1}    
        \and
        Roland Bacon\inst{7}     
        \and
        Fernando Selman\inst{8}  
        \and
        Jarle Brinchmann\inst{5} 
        \and
        Joseph Caruana\inst{1}   
        \and
        Andreas Kelz\inst{1}     
        \and
        Thomas Martinsson\inst{5,9,10}
        \and
        Arlette P\'econtal-Rousset\inst{7}
        \and
        Johan Richard\inst{7}    
        \and
        Martin Wendt\inst{1,6}   
        }
  \institute{Leibniz-Institut f\"ur Astrophysik Potsdam (AIP),                
             An der Sternwarte 16, D-14482 Potsdam, Germany\\
             \email{pweilbacher@aip.de}
         \and
             GEPI, Observatoire de Paris, CNRS, Universit\'e Paris-Diderot,
             Place Jules Janssen, 92190 Meudon, France                        
         \and
             Institut für Astrophysik, Universität G\"ottingen,
             Friedrich-Hund-Platz 1, 37077 G\"ottingen, Germany               
         \and
             ESO, European Southern Observatory, Karl-Schwarzschild Str. 2,
             85748 Garching bei M\"unchen, Germany                            
         \and
             Leiden Observatory, Leiden University, P.O. Box 9513, 2300 RA
             Leiden, The Netherlands                                          
         \and
             Institut für Physik und Astronomie, Universit\"at Potsdam,
             D-14476 Golm, Germany                                            
         \and
             CRAL, Observatoire de Lyon, CNRS, Universit\'e Lyon 1, 9 Avenue
             Charles Andr\'e, F-69561 Saint Genis Laval Cedex, France         
         \and
             ESO, European Southern Observatory, 3107 Alonso de C\'ordova,
             Santiago, Chile                                                  
         \and
             Instituto de Astrof\'{i}sica de Canarias (IAC),                  
             E-38205 La Laguna, Tenerife, Spain
         \and
             Departamento de Astrof\'{i}sica, Universidad de La Laguna,       
             E-38206 La Laguna, Tenerife, Spain
            }
   \date{Received May 13, 2015; accepted XXX YY, 2015}
\abstract{
    We present a new integral-field spectroscopic dataset of the central part
    of the Orion Nebula (M\,42), observed with the MUSE instrument at the ESO
    VLT.
    We reduced the data with the public MUSE pipeline. The output products are
    two FITS cubes with a spatial size of $\sim5\farcm9\times4\farcm9$
    (corresponding to $\sim0.76\,\times\,0.63$\,pc$^2$) and a contiguous wavelength
    coverage of $4595\dots9366$\,\AA, spatially sampled at $0\farcs2$. We
    provide two versions with a sampling of 1.25\,\AA\ and 0.85\,\AA\ in dispersion
    direction. Together with variance cubes these files have a size of 75 and
    110\,GiB on disk. They represent one of the largest integral field mosaics
    to date in terms of information content.
    We make them available for use in the community.
    To validate this dataset, we compare world coordinates, reconstructed
    magnitudes, velocities, and absolute and relative emission line fluxes to
    the literature and find excellent agreement.
    We derive a two-dimensional map of extinction and present de-reddened flux
    maps of several individual emission lines and of diagnostic line ratios.
    We estimate physical properties of the Orion Nebula, using the emission line
    ratios \nii and \siii (for the electron temperature $\Te$) and \sii and \cliii
    (for the electron density $\Ne$), and show two-dimensional images of the
    velocity measured from several bright emission lines.
}
\maketitle

\section{Introduction}\label{sec:intro}
An \hiireg is a diffuse nebula whose gas is heated and ionized by the
ultraviolet radiation of early-type massive stars \citep[see][]{Shi90,OF05}.
\hiiregs are typically found in the arms of spiral galaxies and/or irregular
galaxies and present spectra with strong emission lines visible even at
cosmological distances. Galactic \hiiregs in particular, can be seen as
small-scale versions of the extreme events of star formation occurring in
starburst galaxies \citep[e.\,g.][]{WDF03,AlonsoHerrero09,GarciaMarin09,CCW15}.
As such, they are laboratories offering an invaluable opportunity to study the
interplay between recent and/or on-going star formation -- in particular
massive stars -- and their surrounding interstellar medium, including gas and
dust, at a high level of detail.

One of the best-studied Galactic \hiiregs (and the closest) is the
\object{Orion} Nebula (\object{M\,42}), visible to the naked eye. It is often
one of the first objects targeted with a new instrument; first, to see
if something new can be discovered, and second, to use the plethora of existing
observations for comparison, to validate a new system. A review about the nebula
and its stellar content can be found in \citet{2001ARA&A..39...99O}.
Spectroscopic studies of the ionized gas, confined to one or several slit
positions, have partially characterised the Orion's emission spectrum
\citep[e.\,g.][]{BFM+91,1992ApJ...399..147P,1992ApJ...389..305O,2008ApJ...675..389M,ODH10}.

However, \hiiregs are rarely as simple as the textbook-like
\citet{1939ApJ....89..526S} spheres and Orion is no exception. Indeed, M\,42
is thought to be only a thin blister of ionized gas at the near side of a giant
molecular cloud \citep{1973ApJ...183..863Z,1978A&A....70..769I,2013ApJ...762..101V}.
To make most of the opportunity to observe an \hiireg at the level of detail
offered by Orion, spatially resolved maps with high-quality spectral
information in terms of depth, spatial and spectral resolution are desirable.

Probably,
the most efficient way to gather this information nowadays is the use of
Integral Field Spectroscopy.
\cite{SCV+07} released a first dataset based on this technique to the community
\citep[using the PPak-IFU of PMAS,][]{2006PASP..118..129K}, mapping most of
the Huygens region -- the central part of the nebula with the highest surface
brightness.  However,
the data were taken under non-ideal weather conditions and hence, were poorly
flux-calibrated. They were shallow due to very short exposure time and of low
spatial and spectral resolution. Some of these aspects (depth and spectral
resolution) were improved in a new mosaic mapping a similar area
\citep{2013hsa7.conf..594N}.
However, the spatial resolution of these data were still relatively low. Also,
at the moment, this improved dataset is not publicly available in reduced form.
Additionally, there are several studies with very good data quality in terms of
depth, spectral and spatial resolution, that were devoted to the study of
invidiual targets within the Orion Nebula, and observed the interplay of gas
and stars in proplyds and Herbig-Haro objects
\citep[e.\,g.][]{Vasconcelos05,MesaDelgado11,Tsamis11,MesaDelgado12,Tsamis13,2012MNRAS.421.3399N}.
Therefore, they only mapped very small ($\sim$10\arcsec) rectangular areas.
None of these currently existing datasets satisfy \emph{all} of the following
requirements: i) large mapped area, ii) depth, iii) ample spectral coverage, and
iv) good spatial and spectral resolution.

Here, we present what we call true {\it imaging spectroscopy} of the
Huygens region of the \object{Orion Nebula}, observed with the MUSE
integral-field spectrograph mounted to VLT UT4 ``Yepun''.  MUSE comes close to
producing the ``perfect dataset'' mentioned by \citet{2001ARA&A..39...99O}: it
samples the Huygens region with high spatial sampling (0\farcs2) and reasonable
spectral resolution ($R\sim3000$), and covers a large dynamic range.

Our aim in this work is twofold.
On the one hand, from the technical point of view, this is one of the first
sets of MUSE data and as such, it was taken with the main goal of testing
offsets larger than the field of view and stress-test the data flow system
related to the new instrument.
On the other hand, from a scientific point of view, given the lack of a
high-quality and science-ready set of spectrophotometric data of the whole
Huygens region, we wanted to provide to the community with such data.

In this paper, we first describe the observations and the data reduction
(Sect.~\ref{sec:obsred}), validate the new data against literature values
(Sect.~\ref{sec:fidel}), describe a few unusual artifacts in the MUSE dataset
(Sect.~\ref{sec:artifacts}), and demonstrate how the MUSE datacube can be used
for an analysis of both atomic and ionized gas (Sect.~\ref{sec:warmgas}),
before we conclude with a few general remarks (Sect.~\ref{sec:concl}).

We assume a distance of $D=440$\,pc \citep{2008AJ....136.1566O} for the Orion
nebula. This implies a linear scale of 0.0021\,pc\,arcsec$^{-1}$. The field
of view of the MUSE dataset then corresponds to $\sim0.76\times0.63$\,pc$^2$.

\section{Observations and Data Reduction}\label{sec:obsred}
\begin{figure}
\centering
\includegraphics[width=\linewidth]{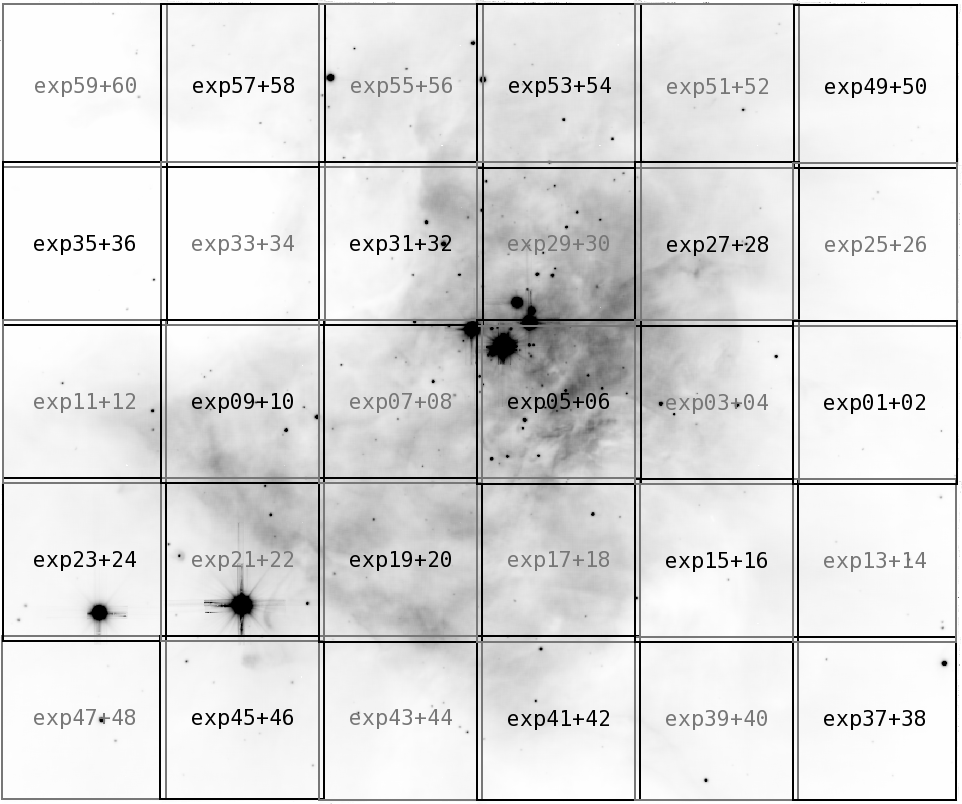}
\caption{Inverse greyscale representation of the white-light image of the final
         mosaic. The positions of the observations are marked and annotated with
         the exposure numbers in the sequence from 1 to 60. Each box represents
         the approximate field of view of a single MUSE exposure, about
         $1\arcmin\times1\arcmin$. The full field covered is
         $\sim5\farcm9\times4\farcm9$, centered on
         $\alpha = 5^\mathrm{h}35^\mathrm{m}17\fs0$,
         $\delta = -5^\circ23\arcmin43\arcsec$, with north to the top and east
         to the left.}
\label{fig:positions}
\end{figure}

M\,42 was observed as part of the first commissioning run
\citep{2014Msngr.157...13B} of the MUSE instrument at the VLT.
After some test exposures during the previous night, a uniform exposure time of
5\,s per exposure was chosen as a compromise to not saturate the bright
emission lines but give sufficient S/N in the outer regions. On 2014-02-16
between 01:02:59 and 03:34:31 (UTC), 60 exposures over a 6-by-5 mosaic were
taken.  Two exposures per position were observed, with the same center but
alternating position angles of 0 and 90 degrees. The positions of the exposures
are schematically shown in Fig.~\ref{fig:positions}. To be able to create a
contiguous grid, the positions were offset by 58\arcsec, which is somewhat
smaller than the MUSE FOV.

Observation of the standard star GD\,71 at 00:15:14 UTC and an airmass of 1.33
allowed a spectrophotometric calibration of the data. No sky exposures were
taken. Daytime calibrations of the morning after the observing night were used.

The observing conditions were good, with photometric sky and DIMM seeing
varying between 0\farcs67 and 1\farcs25.
The M\,42 mosaic was observed after transit, with airmass values ranging from
1.067 to 1.483.  During the observations, the moon had an illumination of 95\%,
a distance of $\sim87^\circ$ from the target, and rose from 10$^\circ$ to
45$^\circ$ in elevation.

The reduction used the dedicated MUSE pipeline \citep[][Weilbacher et al.\
in prep.]{WSU+12} through the {\sc EsoRex} program. We used a development
version of the pipeline, but the code was very close to the 1.0
release\footnote{Available from ESO via
  \url{http://www.eso.org/sci/software/pipelines/muse/muse-pipe-recipes.html}.}
(patches are available on request).

For the basic calibration we followed the standard procedure to reduce MUSE
data: combine 10 bias images to form a master bias, combine 5 lamp-flat
exposures, and use one exposure of each arc lamp to derive the wavelength
solution. 11 skyflats, taken during the evening twilight preceding the science
exposures, were combined and used to create a 3D correction of the illumination
in the range $\lambda = 5000 \dots 8000$\,\AA\footnote{Redder wavelengths were
  excluded from the correction because the 2nd order in the extended mode of
  MUSE created extra artifacts beyond $\sim$8100\,\AA.}.

The geometry of the instrument was derived from a calibration sequence taken on
2014-02-05, the astrometric solution of the MUSE field of view was computed
from an observation of a field in NGC\,3201 on 2014-02-09. These calibrations
were found to be valid for the full period of the first commissioning run of
the instrument and were also shipped with the MUSE pipeline.

We then applied all calibrations to both the standard star exposure and all 60
science exposures, making use of a table of additional bad pixels of the CCDs
that was created after the completion of MUSE commissioning runs, and which is
shipped with the pipeline. For datasets with longer exposure time and lower
contrast, the pipeline usually manages to correct the zero-order of the
dispersion solution on a per-IFU\footnote{{\bf IFU} = {\bf I}ntegral
  {\bf F}ield {\bf U}nit; for MUSE this is one of the 24 subunits consisting of
  image slicer, spectrograph, and CCD, each covering about
  $60\arcsec\times2\farcs5$ on the sky.}
basis using Gaussian fits as centroiding of bright sky emission lines. In the
case of the M\,42 mosaic, extreme contrast differences in some exposures and
IFUs and the low sky emission background made this process unreliable. We
therefore used the following procedure: we assume that the line \oi 5577 is
dominated by telluric emission \citep[see e.\,g.][]{BVV+00}, so it was taken as
baseline reference for each exposure and IFU. Since the shifts on the CCD are
likely smoothly changing during the time of our observation, we iteratively
fitted a linear relation to the {\tt MJD-OBS} against the pipeline-computed
wavelength offset, separately for each IFU. Deviant shifts were aggressively
purged at the $2\sigma$ level.  The wavelength zeropoint was then reset
according to this linear relation with time.

Since the observations were done in extended mode, which incurs a 2nd order
overlap at the red end of the wavelength coverage, the creation of the flux
response curve needed extra care. We ran the pipeline recipe
({\tt muse\_standard}) with both circular flux integration and flux integration
using Moffat profile fits.
Circular apertures were used for wavelengths below 8334\,\AA, as they integrate
slightly more flux than the Moffat fits and create a smoother response curve.
Beyond 8430\,\AA, the circular aperture also integrates significant flux from
the diffuse 2nd order, and the Moffat fit is a better representation of the
total flux in the 1st order. At the transition wavelength, both curves have
approximately the same slope and the curve from the Moffat fit was shifted
slightly to account for the offset of both curves at this wavelength. The
resulting merged response curve was again applied to the data of the standard
star.  A comparison of the reference spectrum with a spectrum extracted from
that calibrated cube showed deviations typically below 5\%.

The merged response and the astrometric calibration were then applied to each
exposure individually. Creating and applying the response function used the
average atmospheric extinction curve by \citet{2011A&A...527A..91P} as shipped
with the MUSE pipeline. We let the pipeline automatically correct the
atmospheric refraction \citep[with the default method using the formula
of][]{1982PASP...94..715F} and the barycentric velocity offset, but did not
attempt to remove the sky background or the telluric absorption in the data.
\footnote{It should be possible to use tools that rely on atmospheric modelling
  instead of dedicated calibration exposures to subtract the sky background and
  remove the telluric features, once spectra are extracted from the cube.
  Examples of such tools are {\sc skycorr} \citep{skycorr,2014ascl.soft08007N}
  and {\sc molecfit} \citep{molecfit,2015ascl.soft01013S}.}
No attempts were made to homogenize the seeing along the wavelength direction,
or between the 60 exposures.

All exposures were combined into a single cube. Since the observation strategy
included rotation, the data were affected by the derotator wobble\footnote{This
  ``wobble'' refers to a decentering of the optical axis of the MUSE derotator
with the axis of the VLT.},
so that each exposure had to be repositioned slightly. Luckily, each pointing
contained at least one star that was also listed in the 2MASS catalog
\citep{2MASS}, so we used the 2MASS positions as reference for the offsets that
were applied when reconstructing the full cube. The effect is that the absolute
astrometry of the final cube is tied to the 2MASS coordinates, similar to the
HST mosaic of the Orion Nebula \citep{2013ApJS..207...10R}.

We created a first full cube with the standard pipeline sampling of
$0\farcs2\times0\farcs2\times1.25\,\AA$.  However, as discussed below, we then
chose a higher wavelength sampling for the final cube, to partially overcome
the undersampling of MUSE data in the dispersion direction.\footnote{In this
  dataset, however, this choice leads to other artifacts, see
  Sect.~\ref{sec:artifacts} and Fig.~\ref{fig:artifacts3}.}
Hence, a second cube was reconstructed with a sampling of
$0\farcs2\times0\farcs2$\,spaxel$^{-1}$ in spatial and a linear step of
$0.85$\,\AA\,pixel$^{-1}$ in wavelength direction. We call this cube the
\fbox{\hr} cube while the cube with the standard sampling is the \fbox{\lr}
cube.  Both cubes have approximately the same wavelength coverage of
$4595\dots9366$\,\AA.  The extent on the sky is exactly the same,
$5\farcm88\times4\farcm92$, but patches up to about 10 spaxels at the edge are
not covered by data and filled with {\tt NaN} values. The total size of the
cubes is $1766\times1476\times3818$ voxels for \lr and $1766\times1476\times5614$
for \hr and they are stored in units of
$10^{-20}\,\mathrm{erg}\,\mathrm{s}^{-1}\,\mathrm{cm}^{-2}\,$\AA$^{-1}$ in the
{\tt DATA} extension of the FITS file. The MUSE pipeline also reconstructs a
variance ($\sigma^2$) cube and stores it in the {\tt STAT} extension in the
same file. Several image extensions are available as well, averaging the cube
either using known filter functions, or using constant weights across
interesting wavelength ranges around some lines (see Table~\ref{tab:extn} for
details). These image extensions are thought to be used only to locate
interesting features in the cube, not for scientific analysis.  The file size
of the full dataset is 75\,GiB (\lr) and 110\,GiB (\hr).

For the purpose of the demonstration in this paper, we finally decided to use
the \lr data for everything except the spatially resolved velocity analysis of
the ionized gas, where \hr gives much lower systematic structures (see
Sect.~\ref{sec:testvelo}). However, the spatial calibration and the
spectrophotometric accuracy is exactly the same for the \hr data, so the values
quoted for the data quality in sections \ref{sec:accwcs}, \ref{sec:magcol},
\ref{sec:absflux}, and \ref{sec:fluxratios} refers to both datasets.

\begin{table}
\caption{FITS extensions in the provided files$^a$}
\label{tab:extn}
\begin{tabular}{l ll}
\hline
EXTNAME               & $\lambda$-range [\AA] & comment \\
\hline
{\bf \verb|DATA|}     & 4595.00\dots9366.05   & data values \\
{\bf \verb|STAT|}     & 4595.00\dots9366.05   & data variance \\
\hline
\verb|white|          & 4650.00\dots9300.00   & \\
\verb|Johnson_V|      &                       & $V$-band filter \\
\verb|Cousins_R|      &                       & $R$-band filter \\
\verb|Cousins_I|      &                       & $I$-band filter \\
\hline
\verb|Halpha|         & 6556.78\dots6568.78   & \ha \\
\verb|NII_both|       & 6542.06\dots6554.06,  & both \nii lines\dots \\
                      & 6577.39\dots6589.39   & \dots (6548 and 6584) \\
\verb|Halpha_NII_OFF| & 6533.05\dots6538.05,  & off-band for \ha\dots \\
                      & 6593.40\dots6598.40   & \dots and \nii \\
\verb|OIII_both|      & 4953.92\dots4965.92,  & both \oiii lines\dots \\
                      & 5000.85\dots5012.85   & \dots (4959 and 5007) \\
\verb|OIII_OFF|       & 4969.92\dots4996.85   & off-band for \oiii \\
\verb|Hbeta|          & 4855.32\dots4867.32   & \hb \\
\verb|Hbeta_OFF|      & 4846.32\dots4851.31,  & off-band\dots \\
                      & 4871.33\dots4876.32   & \dots for \hb \\
\hline
\end{tabular}\\
$^a$ the upper part of the table contains both data cubes, the middle part the
     images from standard pipeline filters, and the bottom images from specially
     created filters
\end{table}

\section{Fidelity of the data}\label{sec:fidel}
Since MUSE is a new instrument and the data reduction software is new, we have
to carefully check the fidelity of the data to ensure its scientific usefulness.

\subsection{Accuracy of the coordinate system}\label{sec:accwcs}
To verify the accuracy of the world coordinate system (WCS) in the MUSE cube,
we determine positions in the reconstructed image integrated over the Johnson
$V$ filter (extension \verb|Johnson_V| in the FITS file). Applying {\sc
daofind} in IRAF\footnote{IRAF is written and supported by the National Optical
  Astronomy Observatories (NOAO) in Tucson, Arizona. NOAO is operated by the
  Association of Universities for Research in Astronomy (AURA), Inc. under
  cooperative agreement with the National Science Foundation.}
to this image yields 259 detections, some of which are spurious sources.

Matching the list of point sources detected in MUSE to the 2MASS catalog \citep{2MASS}
results in 96 matches closer than 1\arcsec. After removing spurious sources,
undetected double stars (listed as source in the 2MASS catalog), and stars
saturated in the MUSE cube, we are left with 90 matched sources. Their
separations are $0\farcs108\pm0\farcs072$ (mean and standard deviation, or
$0\farcs097\pm0\farcs045$ using median and median absolute deviation).
Using the same procedure, but matching MUSE detections against the HST ACS
catalog of the Orion Nebula cluster as given by \citet{2013ApJS..207...10R}, we
find 83 valid matches, giving an overall agreement of $0\farcs163\pm0\farcs078$
(mean and standard deviation).
This is in line with the accuracy of the HST catalog matched against 2MASS
point sources (max. allowed separation 0\farcs5, resulting in
$0\farcs138\pm0\farcs085$) and comparable to the astrometric accuracy of the
2MASS point source catalog itself, $\lesssim100$\,mas given in \citet{2MASS}.

\subsection{Quality of the atmospheric refraction correction}\label{sec:dar}
Since the atmospheric refraction present in the raw MUSE data was corrected by
the pipeline reduction, compact sources in the field do not show strong
gradients across several spatial pixels. Nevertheless, the formula to compute
the refractive index of air \citep[taken from][]{1982PASP...94..715F} is
imperfect, so some residuals are left in the data.

\begin{figure}
\centering
\includegraphics[width=\linewidth]{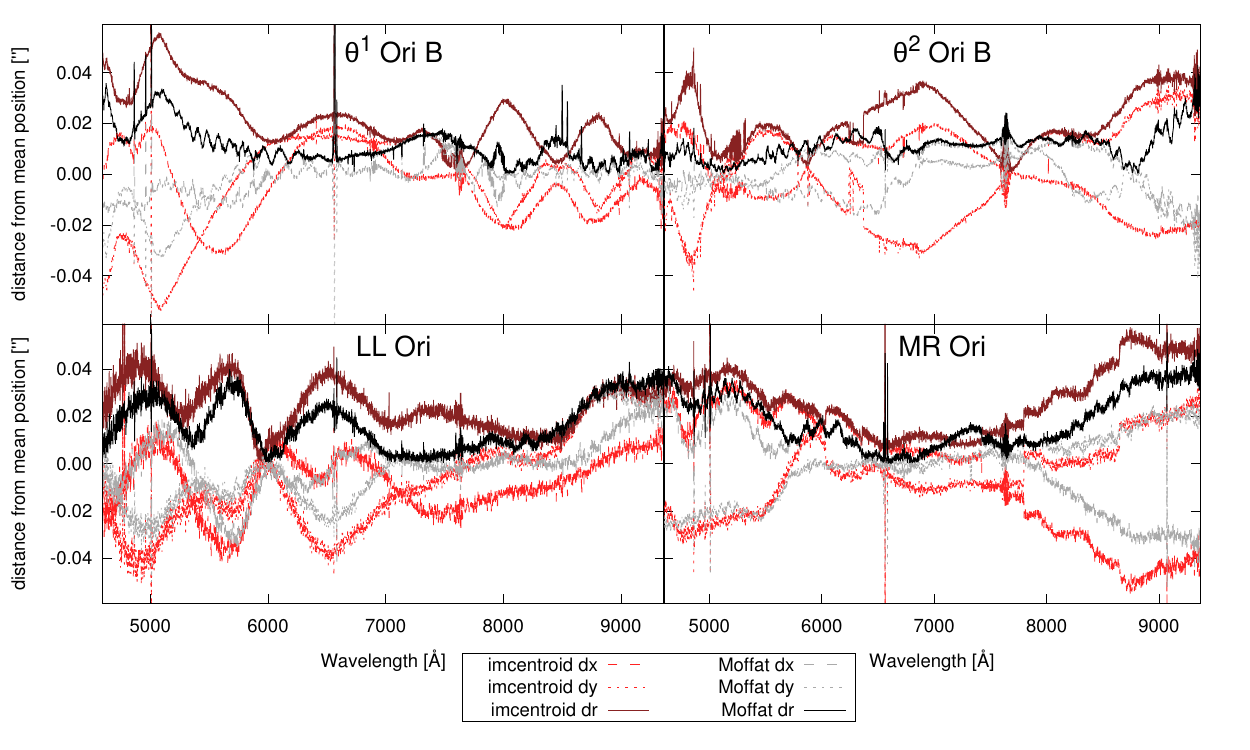}
\caption{Centroiding residuals with wavelength for four bright stars in the
         field. Each panel shows horizontal and vertical distances to the mean
         center of the star in lighter dashed and dotted curves, and the radius
         as solid dark curve. Red curves are for {\sc imcentroid} measurements,
         black lines for Moffat fits.}
\label{fig:dar}
\end{figure}

We test the residuals using the centers of four bright stars in the field.  The
centers of the stars were measured in each wavelength plane of the final cube
by two methods: 1.\ by fitting a Moffat function and 2.\ by computing the
marginal centroid as with {\sc imcentroid} in IRAF. The results are shown in
Fig.~\ref{fig:dar}.

The strongest deviations from the mean centroid position of each star occur in
regions of high background (strong nebular emission lines) and low throughput
(telluric absorption). Ignoring these wavelength ranges, the typical deviations
from the mean centroid position are smaller than 0\farcs05 or 1/4th of a
spatial element of the cube.

\subsection{Magnitudes and colors}\label{sec:magcol}
We used the same stellar spectra already extracted using the Moffat fit in
Sect.~\ref{sec:dar} to determine how well we can reproduce stellar magnitudes
using the MUSE data.

\begin{figure}
\centering
\includegraphics[width=\linewidth]{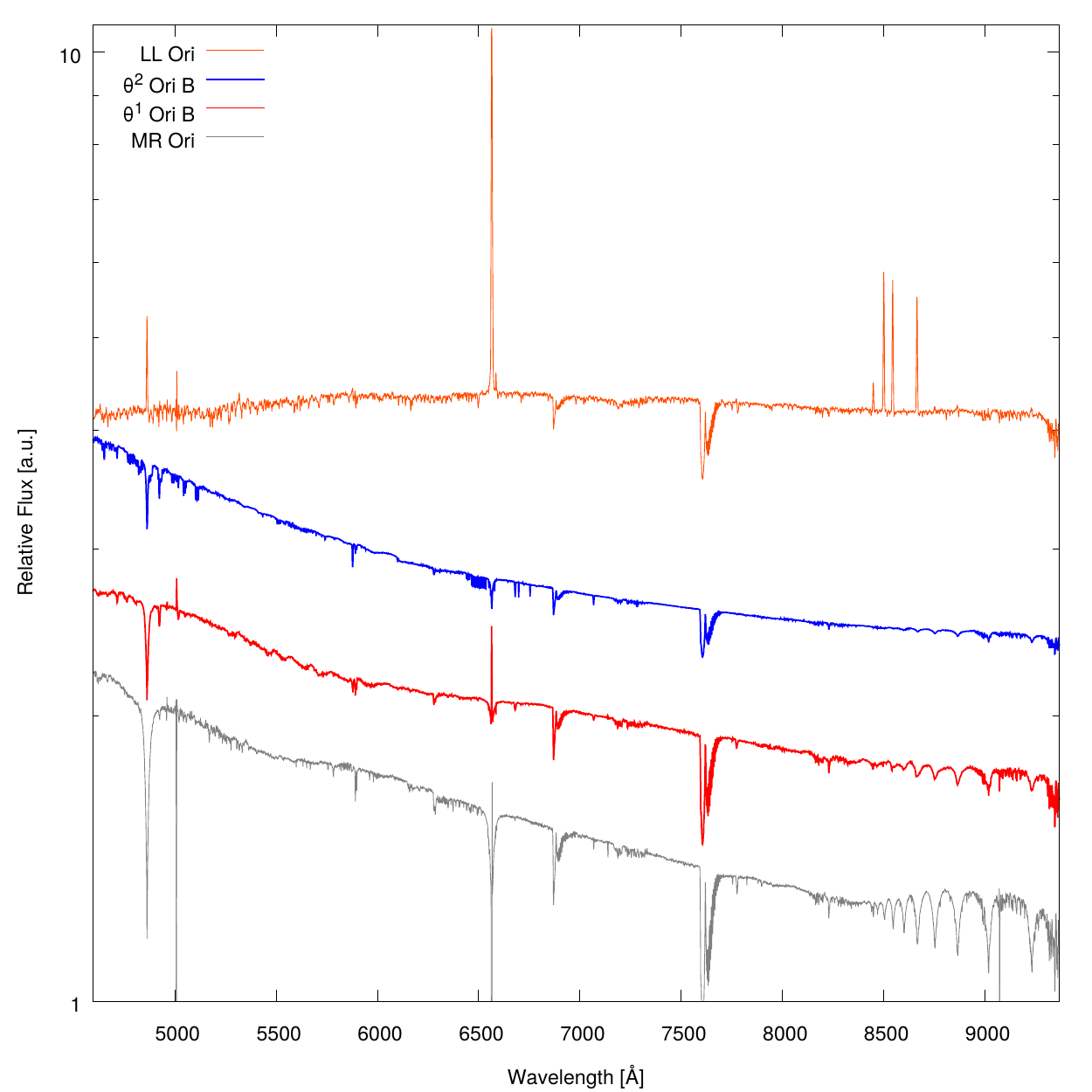}
\caption{Extracted spectra of four bright stars in the field; the flux is
         displayed in arbitrary logarithmic units, the individual spectra
         are scaled and offset for better visibility, and not in order of
         stellar magnitude.}
\label{fig:stellarspectra}
\end{figure}

The extracted spectra are shown in Fig.~\ref{fig:stellarspectra}. The telluric
absorption that is visible does not significantly affect measurement of the
integrated magnitudes ($R$-band: 0.007\,mag, $I$-band: 0.023\,mag).
Nevertheless, for this test, we replaced the absorbed regions in the spectrum
with an average value before integrating the spectra over the filter functions
(we used Johnson $V$, and Cousins $R$ and $I$).

The effect of nebular line emission that was not optimally subtracted by the
Moffat fit is less certain. Indeed, \object{LL\,Ori} shows intrinsic Balmer and
CaII-triplet emission that cannot be explained by residuals of nebular
emission \citep[see also][]{1997AJ....113.1733H} whereas small residuals of
\oiii 5007 and \nii 6584 are very likely of nebular origin. Since the
equivalent widths of the emission features are low --
$\mathrm{EW}(\mathrm{H}\alpha)\approx-42.5$\,\AA\ for LL\,Ori -- the influence
on the broad-band magnitudes is rather low.

\begin{table}
\caption{Comparison of integrated magnitudes to the literature.}
\label{tab:mags}
\begin{tabular}{l crc}
\hline
object                & band & $m_\mathrm{MUSE}$ [mag] & $m_\mathrm{lit} - m_\mathrm{MUSE}$ [mag] \\
\hline
\object{$\theta^1$\,Ori\,B} & $V$ &         7.990      & -0.03$^1$ \\
                      & $R$  &              7.760      & \\
                      & $I$  &              7.422      & \\
\object{$\theta^2$\,Ori\,B} & $V$ &         6.389      & 0.009$^1$, 0.021$^4$ \\
                      & $R$  &              6.403      & -0.104$^1$ \\
                      & $I$  &              6.356      & (4.694$^{2,3}$), 0.054$^4$ \\
\object{LL\,Ori}      & $V$  &             11.499      & 0.018$^2$, 0.021$^4$ \\
                      & $R$  &             10.835      & \\
                      & $I$  &             10.218      & -0.184$^2$,
                                                         -0.038$^4$ \\
\object{MR\,Ori}      & $V$  &             10.551      & 0.027$^2$ \\
                      & $R$  &             10.286      & \\
                      & $I$  &             10.010      & -0.174$^2$ \\
\hline
\end{tabular}\\
$^1$ compared to magnitude given by \citet{2002yCat.2237....0D}\\
$^2$ compared to magnitude given by \citet{2009ApJS..183..261D}\\
$^3$ the value given by \citet{2009ApJS..183..261D} for \object{$\theta^2$ Ori B}
     ($m_I=11.050$\,mag) is clearly faulty\\
$^4$ compared to magnitudes given by \citet{1997AJ....113.1733H}
\end{table}

An agreement of a few hundreths of a magnitude in the $V$-band can be regarded
as very satisfactory. The $V$-band differences quoted in Tab.~\ref{tab:mags}
correspond to flux differences of up to 2.7\%.

However, the comparison with $R$ and $I$ filters, where available, remains
puzzling. Only the $I$-band magnitudes given in the table of
\citet{1997AJ....113.1733H} are close to our measurements, and these are only
available for two of our four stars used for this experiment.  As we have
argued in Sect.~\ref{sec:obsred} and checked in a different way in
Sect.~\ref{sec:fluxratios}, differences of 0.1 to 0.18\,mag or up to 18\% in
flux \citep[as seen relative to the measurements
of][]{2002yCat.2237....0D,2009ApJS..183..261D} are unlikely to be a problem
with the relative flux calibration of the MUSE spectra, which is accurate to at
least 5\%.

To further investigate the difference, we also convolved our spectra with the
filter plus CCD throughputs of the ESO WFI setup used by
\citet{2009ApJS..183..261D}. This made the agreement even worse.  Since all
stars in our field of view are likely variable at some level -- of the four
stars we analyzed here, all except $\theta^2$\,Ori\,B are listed in the
variable star catalog of \citep{2009yCat....102025S} -- one should not expect
high precision of the comparison. But as variability likely affects
observations in different filters in a similar way, this cannot explain the
differences we see here.
We therefore have to assume that the zeropoints applied by Da Rio et al.\
include an unknown component that causes a shift of the central effective
wavelength of the red filters, but less of a shift for the green
filter.\footnote{This could be due to the relative throughput of atmosphere or
  telescope that are unknown to us, see e.\,g.\ \citet{2010AJ....139.1628D} for
  details on filter profile determinations and the effect of the atmosphere.}
Since the Orion Nebula is too bright for SDSS stellar photometry to work and all
four stars are marked as saturated in the HST ACS catalog of
\citet{2013ApJS..207...10R}, we cannot cross-check our reconstructed magnitudes
with a better-studied photometric system.

\subsection{Derived velocities}\label{sec:testvelo}
MUSE has a moderate velocity resolution (about 107\kms at 7000\,\AA), and the
line spread function is slightly undersampled.\footnote{MUSE has a typical FWHM
  of 2.5\,\AA\ sampled at about 1.25\,\AA\,pixel$^{-1}$.}
As a consequence it is challenging to measure accurate velocity centroids for
single narrow spectral features such as emission lines in HII regions.
Nevertheless, we compare our derived velocities against the values given by
\citet[][hereafter B00]{BVV+00}.

The MUSE cube is corrected to barycentric velocities\footnote{The difference
  between barycentric and heliocentric velocities at the time of observations
  was less than 10\,m\,s$^{-1}$.},
so we can directly check our velocities against their heliocentric reference
value of $+11.9$\kms. If we extract a spectrum over the same aperture as the
``blue'' slit of B00, and fit all bright and a few fainter emission lines with
Gaussian profiles, we can plot the resulting velocities for both of our cubes
(\lr and \hr) as well as the reference values from B00.

\begin{figure}
\centering
\includegraphics[width=\linewidth]{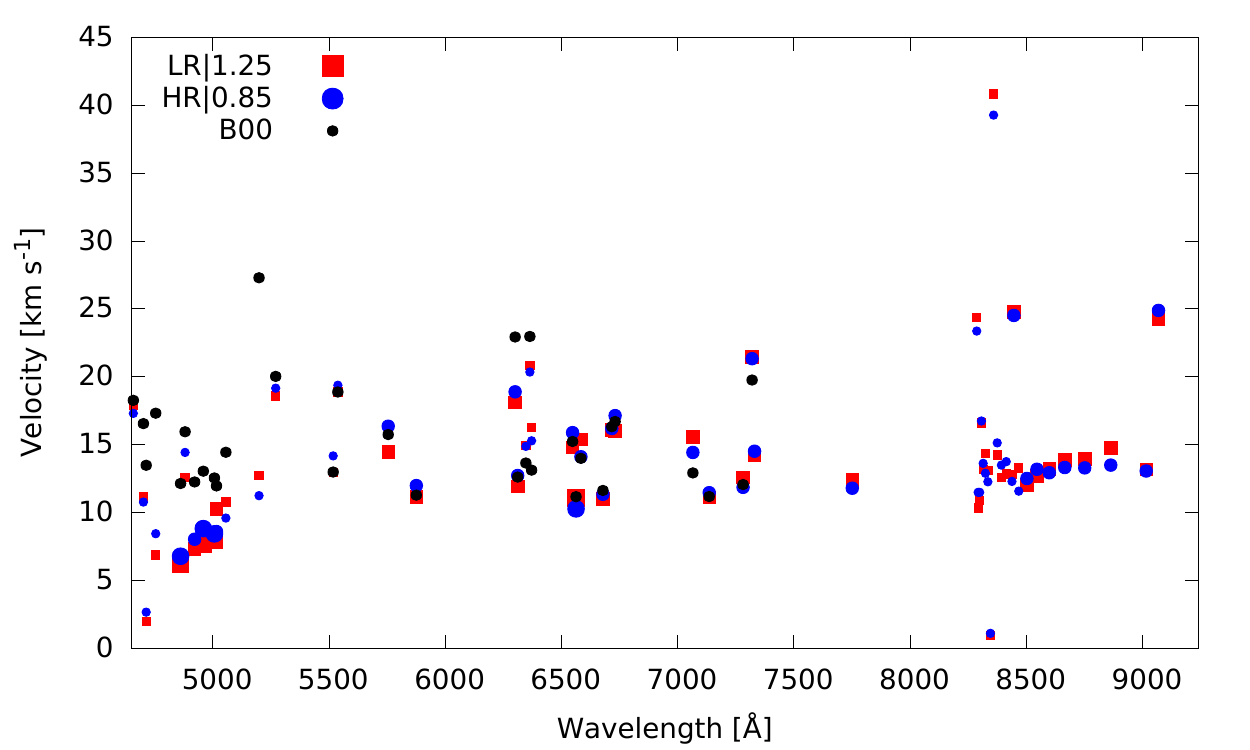}
\caption{Velocities measured within the ``blue'' slit of \citet[][B00]{BVV+00}.
         These are corrected to barycentric velocity for our data (red squares
         and blue circles) and to heliocentric for B00 (black circles). The
         point size of our measurements changes with the relative flux of the
         lines, such that the brightest lines have the largest symbols.}
\label{fig:velos_lines}
\end{figure}

In Fig.~\ref{fig:velos_lines} we show the result. We can reproduce the
average velocity of about 12\,\kms reported by B00.\footnote{Note that we
  consistently use reference wavelengths from the NIST database \citep{NIST_ASD_2014}
  for this plot, e.\,g.\ 6562.819\,\AA\ for \ha and 9229.014\,\AA\ for Pa9.}
This is seen for most of the Paschen lines in the very red (that were not
measured in B00's setup) and -- with some scatter -- for most other lines in
the wavelength range $5250 < \lambda < 8000\,\AA$. The bright lines below
5250\,\AA, however, show a deviation from this mean velocity, in the sense that
we measure velocities about 6\,\kms below those derived by B00. While the
fainter lines at similar wavelengths do not all show this offset, they are
partly blended with neighboring lines, and are therefore less trustworthy.

Several strong outliers are visible at the red end of the spectrum in
Fig.~\ref{fig:velos_lines}. These are three of the fainter Paschen lines (Pa22, Pa23,
and Pa30) and most likely due to blending with another unidentified weak line.
However, the strong and relatively isolated lines OI 8446 and \siii 9069 also
show a strong offset of $\sim12$\,\kms. Since the surrounding Paschen lines
follow the normal trend very well, this casts doubts on the reliability of the
reference wavelengths \citep[we used 8446.462 and 9068.6 from the NIST database][]{NIST_ASD_2014}.
Indeed, taking the reference value of 9068.9\,\AA\ as quoted by
\citet{1992ApJ...389..305O} for \siii, we find a velocity of
14.35 and 14.97\,\kms for our \lr and \hr data, respectively, perfectly in line
with the general trend.

We therefore conclude that in the range $\lambda < 5250\,\AA$ the MUSE data
likely shows a problem with the wavelength calibration, on the level of up to
0.1\,\AA\ (less than 1/10th of a MUSE pixel), while no systematic deviations
larger than $\sim3$\,\kms were found for wavelengths $\lambda > 5250\,\AA$.

\begin{figure}
\centering
\includegraphics[width=\linewidth]{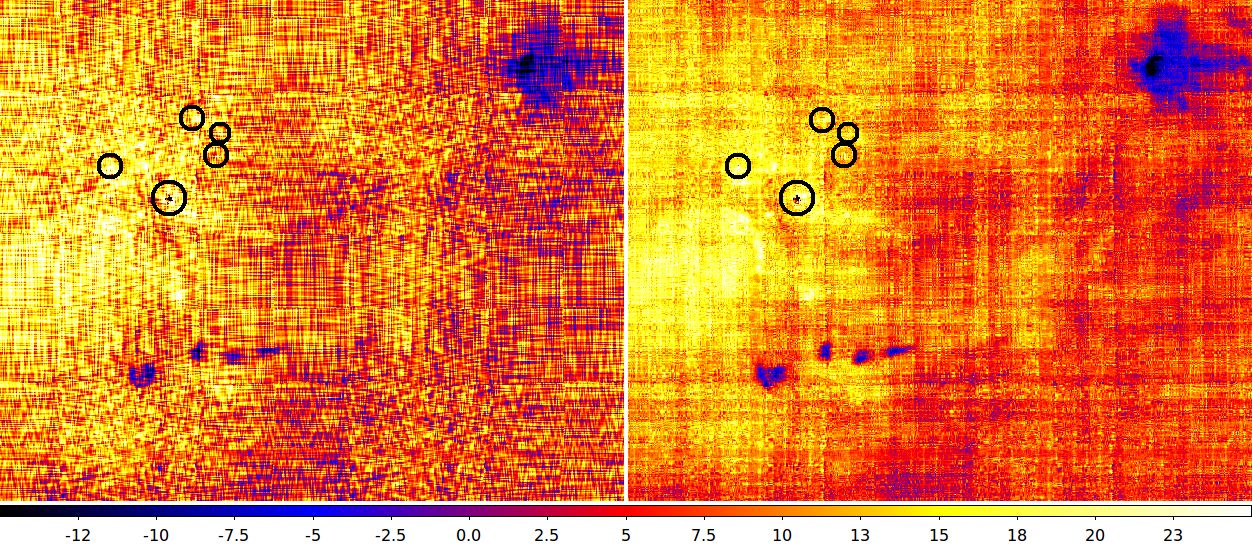}
\caption{Velocity field as traced by the line \oiii 5007, in part of the MUSE
         field of view.
         {\bf Left}: using the \lr cube; {\bf right}: using the \hr cube.
         The linear color scale is $v_\mathrm{bary}=-15\dots25$\kms, the
         brightest stars of the Trapezium cluster are marked as black circles.}
\label{fig:OIII_velos_part}
\end{figure}

We then compare maps of emission line velocities over the full field. These
velocities were derived using single Gaussian profile fits to individual lines,
and the maps show systematic patterns. These systematics are much more
pronounced in the \lr cube than in the \hr data. This is demonstrated in
Fig.~\ref{fig:OIII_velos_part}, which shows the two-dimensional velocity field
recovered from Gaussian centroids of the \oiii 5007 emission line, for a part
of the field around the Trapezium stars.  The strong
horizontal and vertical stripes show the influence of the per-slice sampling
in the MUSE field of view on the derived velocity field. This effect is much
reduced in the \hr cube, where the higher sampling of the line profile allows
for more stable line fits. Only a residual pattern of the IFU structure is still
visible (about 12 pixel wide stripes).
For derivation of spatial velocity fields, it is therefore recommended to use
the \hr cube.

To determine if the blue wavelength calibration problem discussed above is an
absolute offset with wavelength or has gradients across the field, we also
compute average and standard deviation of velocity difference maps for a few
lines: we find $v(\mathrm{H}\alpha) = 14.5\pm5.9$\kms (\lr) and
$14.4\pm4.5$\kms (\hr), so that the overall velocity field in \ha is very
comparable between both datacubes. Taking the statistics from the difference
map of \ha and \pazeta, we find $v(\mathrm{H}\alpha) - v(\mathrm{Pa}\zeta)=
-1.1\pm9.6$\kms (\lr) and $-1.0\pm8.3$\kms (\hr), i.\,e.\ a very good agreement
between velocities derived from \ha and the Paschen lines across the whole
field of view. Comparing \ha and \hb in the same way, we find a similar offset
as discussed above for one slit position: $v(\mathrm{H}\alpha) -
v(\mathrm{H}\beta) = 4.9\pm10.5$\kms (\lr) and $4.9\pm9.4$\kms (\hr). We
therefore conclude that the relative wavelength calibration across the field is
very stable when measuring each emission line individually.

We also briefly compare the map shown in Fig.~\ref{fig:OIII_velos_part} with
the literature.  The velocities we measure are at odds with those measured by
\citet{2001AJ....122.1928R} using an optical Fabry-P\'erot instrument, who find
strongly blueshifted velocites (up to and around $-100$\kms) over much of the
nebula, in the H$\alpha$ line. In the vicinity of HH\,203 they also find
velocities of around $-75$\kms in the H$\alpha$ line, whereas we see $v \approx
-20$\kms at the most blueshifted part of the same jet (also see
Fig.~\ref{fig:velos_full}). Our measurements are more comparable to the
velocities measured in the near-infrared by \citet[][with a different
instrument also called MUSE]{2002ApJ...566..910T}, who see approximately
$0$\kms in the region around the Trapezium stars, and $\sim -20$\kms in HH\,203
in \hei\,10830 and approximately $-15$\kms in Pa$\beta$.

Since our data are of lower spectral resolution than some other studies, we
should be sensitive only to high-velocity features in the {\em primary}
component of each emission line.  Since strong bipolarities as reported by
e.\,g.\ \citet{2004AJ....127.3456D} around some proplyds are only detectable in
fainter components -- which in our data are blended with the main line profile
-- we cannot detect such features.

\subsection{Absolute fluxes}\label{sec:absflux}
\begin{figure}
\centering
\includegraphics[width=\linewidth]{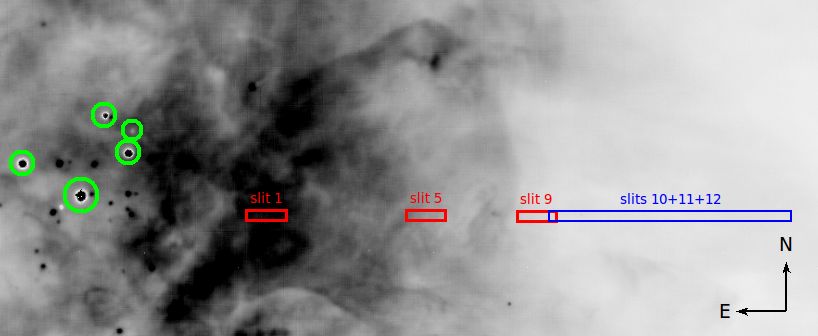}
\caption{Literature slit positions taken from \citet{BFM+91} on top of a
         \hb flux map. The compass has an extent of 10\arcsec, the
         Trapezium stars are marked with green circles.}
\label{fig:slitpos}
\end{figure}

There are only few publications on the Orion Nebula which also quote absolute
emission line fluxes; most rely on flux ratios.
To check the flux calibration of our data, we extracted spectra at positions
given in the literature. For this purpose, we used the regions functionality of
DS9 which allowed easy, interactive placement of rectangular boxes onto the WCS
positions, approximating the given slit lengths and widths. Then, the ``3D plot''
function was used to create spectra summed over these regions.

Following the example of \citet[][hereafter ODH10]{ODH10}, we first extracted a
96\farcs8 box corresponding to the corrected position (4\arcsec\ south of
$\theta^1$\,Ori\,C as described in ODH10) for slits 10+11+12 of
\citet[][hereafter B91]{BFM+91}. The positions of the slits reproduced on top
of the MUSE data are displayed in Fig.~\ref{fig:slitpos}. We then measured the
principal emission lines for comparison using Gaussian fits with {\sc splot} in
IRAF\footnote{A Voigt profile gives better fits to the wings of the lines, but
  the integrated fluxes are typically only $\lesssim1\%$ larger.  For
  simplicity and consistency with other parts of this paper, we therefore use
  Gaussian profiles.}
We find $F_\mathrm{H\beta} = 1.623\times10^{-13}\,\mathrm{erg}\,\mathrm{s}^{-1}\,\mathrm{cm}^{-2}\,\mathrm{arcsec}^{-2}$,
i.e.\ a 6\% difference to B91 ($1.73\times10^{-13}$ but 14.6\%
difference to the measurement of $1.90\times10^{-13}$ by ODH10).

\begin{table}
\caption{Comparison of \hb flux measurements with the literature.}
\label{tab:hbeta}
\begin{tabular}{l ccc}
\hline
slit                  & $F_\mathrm{H\beta,literature}$ & $F_\mathrm{H\beta,MUSE}$ & rel.\ change \\
\hline\hline
B91$^a$ slit 10+11+12 & $1.73\times10^{-13}$           & $1.623\times10^{-13}$    &   6.2\% \\
c.f.~ODH10$^b$        & $1.90\times10^{-13}$           & $1.623\times10^{-13}$    &  14.6\% \\
\hline
B91 slit 1            & $9.99\times10^{-13}$           & $1.229\times10^{-12}$    & -23.0\% \\
B91 slit 5            & $7.39\times10^{-13}$           & $5.314\times10^{-13}$    &  28.1\% \\
B91 slit 9            & $2.22\times10^{-13}$           & $2.210\times10^{-13}$    &   0.5\% \\
\hline\hline
B91 slit 9, 2\arcsec\ E\ $^{c}$ & $2.22\times10^{-13}$ & $2.532\times10^{-13}$    & -14.1\% \\
B91 slit 9, 2\arcsec\ W\ $^{c}$ & $2.22\times10^{-13}$ & $2.064\times10^{-13}$    &   7.0\% \\
B91 slit 9, 2\arcsec\ N\ $^{c}$ & $2.22\times10^{-13}$ & $2.136\times10^{-13}$    &   3.8\% \\
B91 slit 9, 2\arcsec\ S\ $^{c}$ & $2.22\times10^{-13}$ & $2.273\times10^{-13}$    &  -2.4\% \\
\hline\hline
\end{tabular}
$^a$ B91 = \citet{BFM+91}\\
$^b$ observations by \citet{ODH10} at position of B91\\
$^c$ value measured by B91 in their slit 5 as reference
\end{table}

We carried out the same comparison for slits 1, 5, and 9 of B91, similarly
correcting the position in declination. The results can be found in
Tab.~\ref{tab:hbeta}. The differences in absolute flux measurement are up to
28\%, unexpectedly high, if the positions would be recovered perfectly.
However, the discussion in ODH10 shows that the positions and widths used
during longslit spectroscopic observations are rather uncertain, errors up to a
few arc seconds may occur. We therefore briefly investigate what the effect of
small shifts applied to the best recovered position have on the absolute fluxes
of the \hb line.  Applying 2\arcsec\ (the width of the B91 slit) offsets
to the position of slit 9 (the one that best recovers the flux determined by
B91) in all 4 directions shows that flux difference of up to $\sim$15\% can
easily occur.

We therefore conclude that the main cause of the differences in absolute \hb
flux with respect to literature values are likely caused by uncertainties in
the slit positions we can recover.

\subsection{Flux ratios}\label{sec:fluxratios}
Flux ratios, relative to \hb or \hei 6678 are given in many publications on the
Orion Nebula. However, it is not always straightforward to reproduce the slit
placement accurately enough to derive a meaningful comparison. E.\,g.\ the
observations carried out by \citet{1992ApJ...389..305O} and \citet{BVV+00} were
located in regions of strong emission-line gradients. Additional problems, like
the unknown effect of atmospheric refraction on the literature line ratios make
the comparison even more difficult.

\begin{table}
\caption{Comparison of emission line ratios to the literature.}
\label{tab:relfluxes}
\begin{tabular}{r@{\hspace{5pt}}l | l@{\hspace{5pt}}l@{\hspace{5pt}}l | r@{}r}
$\lambda$& ID       &   & $F/F_\mathrm{H\beta}$   &
                                        & rel.   & change    \\
  \ [\AA]&          &B91$^a$&ODH10$^b$&MUSE$^c$&$\frac{\mathrm{B91}}{\mathrm{MUSE}}$&$\frac{\mathrm{ODH10}}{\mathrm{MUSE}}$ \\
\hline
4861.48 & \hb   & 1.00  & 1.00  & 1.000 &  0.0\% &  0.0\% \\
4959.09 & \oiii & 0.92  & 0.93  & 0.934 & -1.5\% & -0.4\% \\
5007.03 & \oiii & 2.76  & 2.75  & 2.813 & -1.9\% & -2.3\% \\
6563.08 & \ha   & 3.34  & 3.20  & 3.358 & -0.5\% & -4.9\% \\
6716.81 & \sii  & 0.061 & 0.058 & 0.057 &  5.9\% &  1.0\% \\
6731.23 & \sii  & 0.076 & 0.068 & 0.071 &  6.5\% & -4.5\% \\
7065.56 & \hei  & 0.047 & 0.048 & 0.049 & -4.8\% & -2.6\% \\
\hline
\end{tabular}\\
$^a$ flux ratio from \citet{BFM+91} as quoted by ODH10\\
$^b$ flux ratio from \citet{ODH10} (their Tab.~1)\\
$^c$ flux measured in MUSE data in the same area using Gaussian line fits
\end{table}

Finally, we reproduced the approach and slit placement of
\citet[][ODH10]{ODH10}, which has the advantage of being located in an area
with shallower gradients. This also lets us compare again the relevant lines
with \citet[][B91]{BFM+91}.

We used our extracted spectrum and the measurement procedure that we
already discussed in Sect.~\ref{sec:absflux}.  The result is presented in
Tab.~\ref{tab:relfluxes}, as fluxes relative to the \hb line. The MUSE
result lies approximately between the measurements of B91 and ODH10, with
maximum deviations of up to 6.5\% to one of the references, with a maximum
deviation of 3.7\% to the mean of both reference measurements. This is the
deviation expected given the flux calibration accuracy quoted in
Sect.~\ref{sec:obsred}.

Since the night was photometric, we make no attempts to check other regions of
our cube, but assume that the relative fluxes are accurate to $\sim5$\% over
the full field.

\section{Artifacts visible in the data}\label{sec:artifacts}
Potential users of the data should be aware of a few artifacts present in the
data. Some of them are due to the imperfect calibrations taken at the time of
the first commissioning with the MUSE instrument, others are instrumental
features that are hard to model and hence cannot be removed. These effects
will be described in detail by Bacon et al.\ (in prep.).

\begin{figure}
\centering
\includegraphics[width=\linewidth]{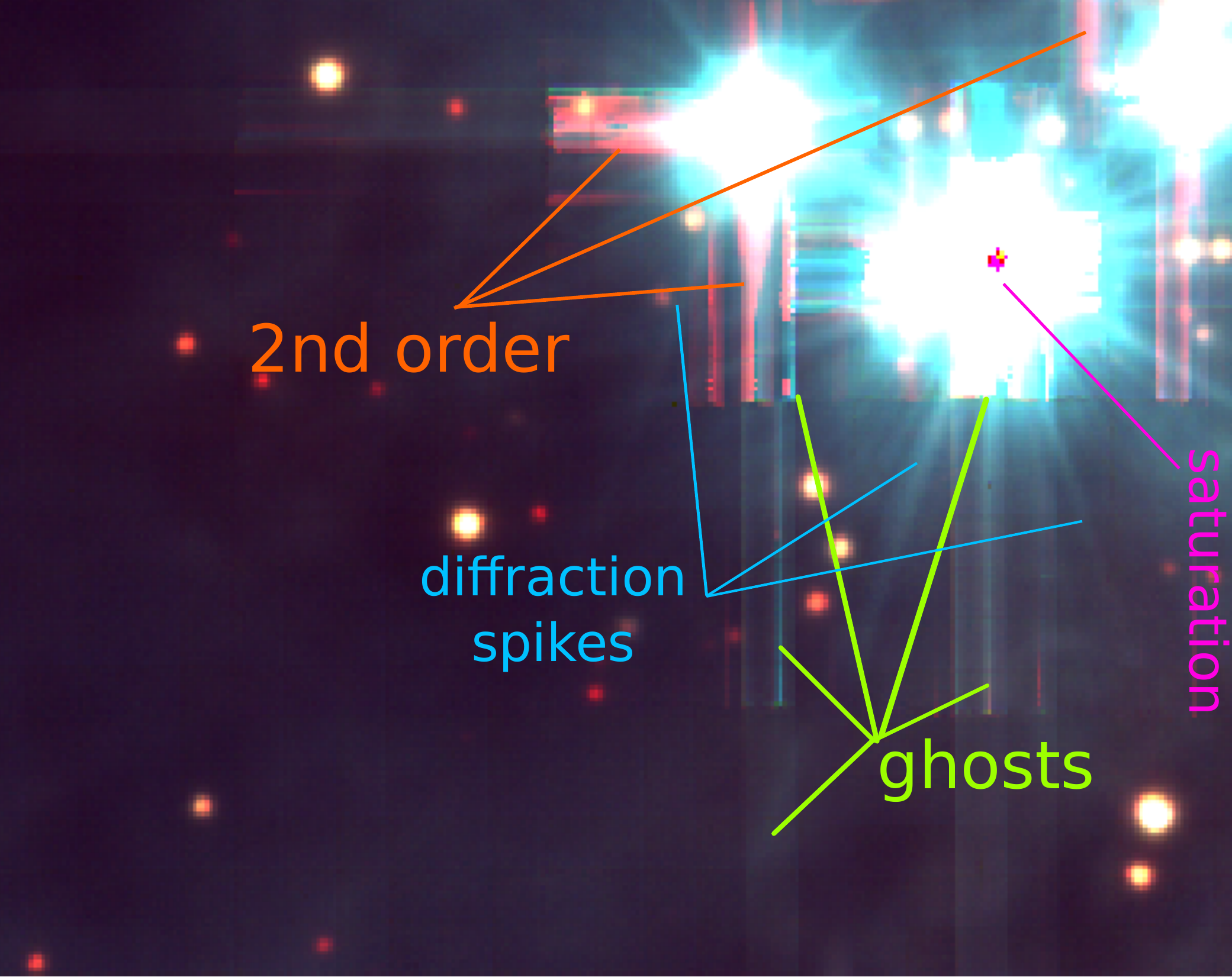}
\caption{Color composite of the region near the Trapezium stars. The colors
         are created from continuum regions of the spectrum, with
         blue: 4600\dots4800\,\AA, green: 5300\dots5500\,\AA, and
         red: 9100\dots9200\,\AA.
         Ghosts are visible as blueish vertical and horizontal stripes,
         second order as reddish striping, and saturation as magenta pixels.}
\label{fig:artifacts1}
\end{figure}

The brightest part of the data, the region around the Trapezium is particularly
affected. Fig.~\ref{fig:artifacts1} shows this region in a color composite,
highlighting the continuum in three wavelength bands.

At all wavelengths, {\it ghosts} around bright objects are visible as vertical
and horizontal stripes. These are likely due to internal reflections in the
spectrographs, resulting in a faint background (at the level of $6\times10^{-6}$,
Bacon et al.\ in prep.) across the CCD, mostly affecting the slice in which the
bright object is located but also neighboring slices. The cross-pattern this
creates is due to the observations with the two position angles of 0 and 90
degree.
Note that these ghosts are not the same as the usual diffraction spikes seen
in pure imaging data. These exist in the MUSE data as well (see
Fig.~\ref{fig:artifacts1}), but are fainter than the ghosts, and smoothly
decrease with radial distance from the star.

\begin{figure}
\centering
\includegraphics[width=\linewidth]{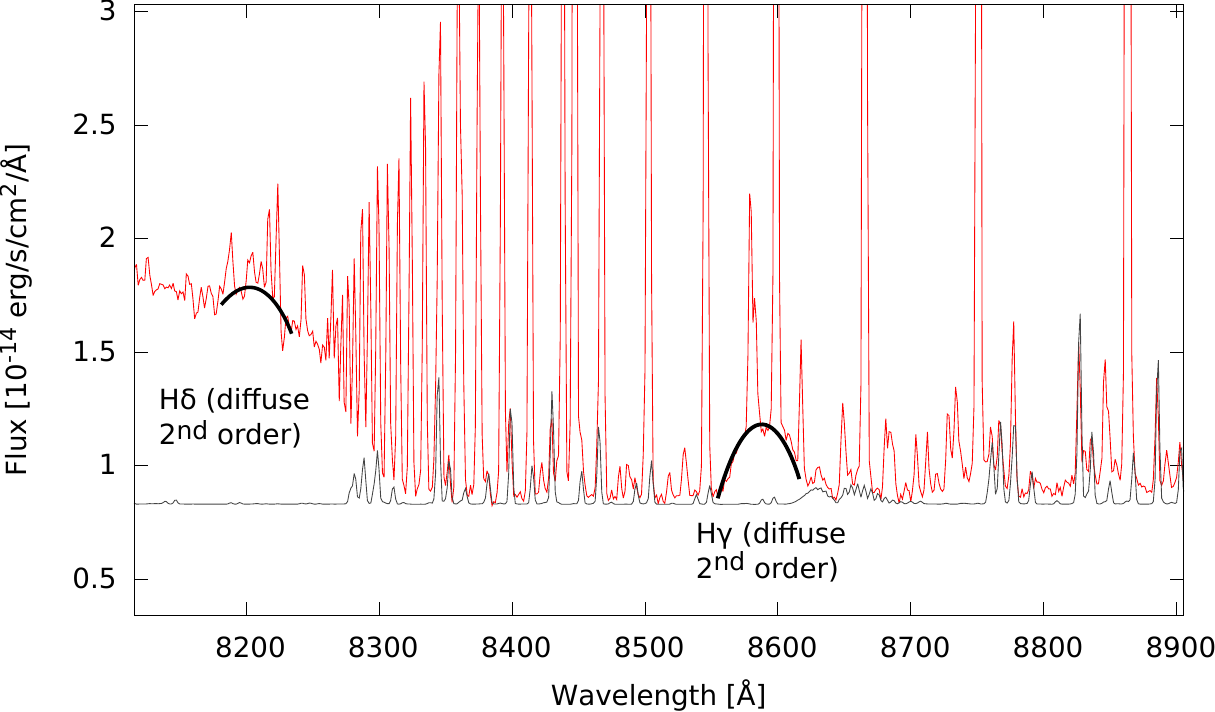}
\caption{Spectral region around the Paschen limit. The red spectrum was
         extracted from the MUSE cube at the approximate position of slit1 of
         \citet{BFM+91}. In grey, a typical Paranal sky spectrum is shown.
         Broad second order contamination by prominent blue Balmer emission
         lines is marked in black.}
\label{fig:artifacts2}
\end{figure}

In the very red part of the spectrum, the {\it second spectral order} also
becomes apparent. Since this order is unfocused in MUSE, it appears broadened compared
to real features in the data.  In Fig.~\ref{fig:artifacts1} the 2nd order can
be seen spatially, as red striping, again in horizontal and vertical
directions.
The second order is also visible in spectral direction, as displayed in
Fig.~\ref{fig:artifacts2}\footnote{Note that the 2nd order is not wavelength
  calibrated well, nor does it appear exactly on the same horizontal (or
  spatial) position on the CCD. This likely causes the bumps to appear somewhat
  offset from the expected position of $2\times\lambda_1$.}.
The nebular continuum is almost featureless, so this merely creates an additional
offset, but strong emission lines in the blue can create broad bumps in the
spectrum. However, if care is taken to locally fit or subtract the background when
integrating line fluxes, these broad bumps should not cause any problems when
extracting signal from the cube.

The brightest stars are also {\it saturated} near their peak. Part of this
saturation appears as magenta pixels in Fig.~\ref{fig:artifacts1}. However,
around the strongly saturated pixels, a few more pixels may be strongly
negative, or have positive values below the real value. No attempts were made
to mask these out.

Since neither {\it telluric emission} (Fig.~\ref{fig:artifacts2}) nor {\it
absorption} (Fig.~\ref{fig:fullspectrum}) were treated in our reduction, both
still show up in extracted spectra. The nebular emission towards the Huygens
region is very bright compared to the sky, most measurements are unaffected by
more than a few percent. Notable exceptions are emission lines that are of
comparable brightness to the sky background, such as \oi 5577 or -- in a
few places where the nebula is faint -- \oi 6300.

\begin{figure}
\centering
\includegraphics[width=\linewidth]{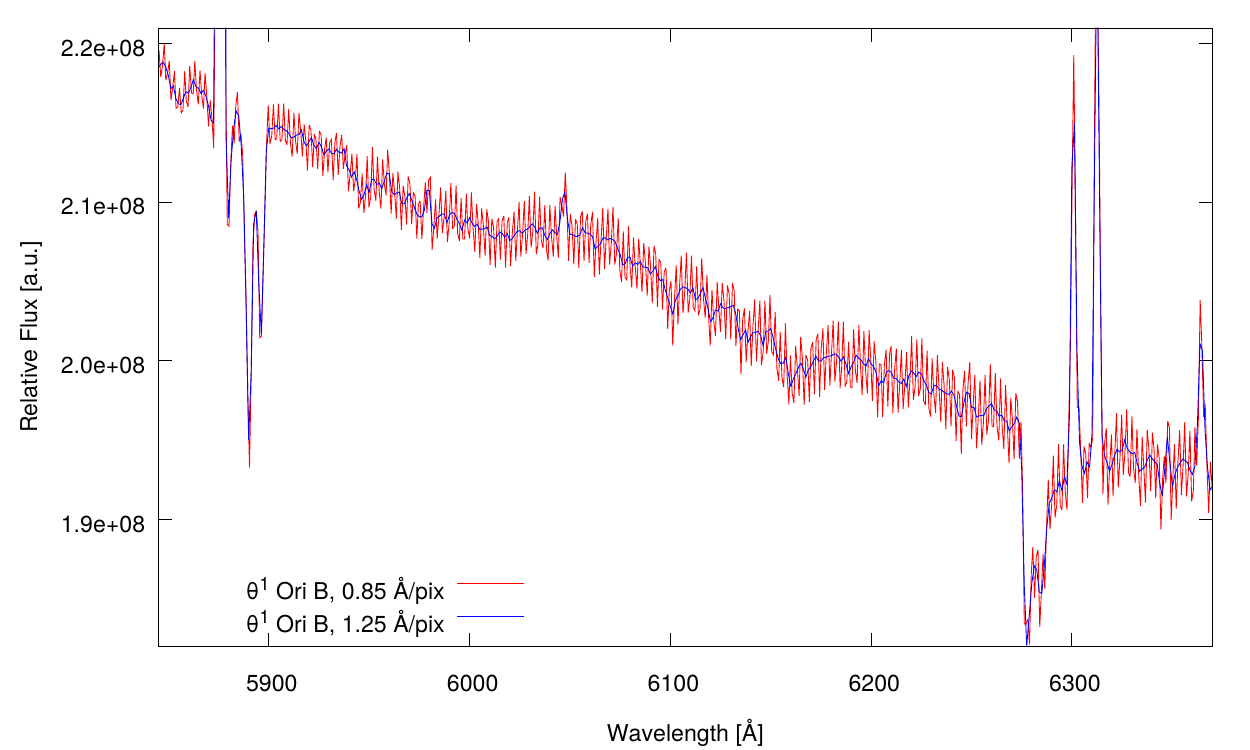}
\caption{``Wiggles'' in the extracted \hr spectrum of $\theta^1$\,Ori\,B,
         compared with normal \lr spectrum.}
\label{fig:artifacts3}
\end{figure}

When extracting spectra of (bright) point sources from the \hr MUSE cube, {\it
wiggles in the continuum} can be seen (Fig.~\ref{fig:artifacts3}). The origin
of this feature is not fully understood, but it is likely caused by a change of
the spatial point-spread function (PSF), i.\,e.\ variations in seeing between
exposures.
Since the two exposures per position are interleaved in the final \hr cube and
-- depending on the exact sampling in the cube -- often contribute to
alternating wavelength planes, any variations in seeing between exposures
effectively causes the PSF to also vary between adjacent planes in the cube.
When integrating the flux of a star in an aperture that is too small to integrate
the whole flux in both seeing conditions, the output spectrum looks particularly
`wiggly', as displayed in Fig.~\ref{fig:artifacts3}.
The problem is mitigated when using PSF-matched spectral extraction
\citep{KWR13} where the PSF is fitted to each wavelength plane separately
instead of using simple aperture extraction. We confirmed this by testing
extraction strategies in a different MUSE two-exposure dataset (of a globular
cluster, Kamann et al., in prep.) at \lr and \hr output sampling. However,
since the PSF fit is imperfect, even then the wiggles do not completely
disappear from \hr data.  For stellar work, we therefore recommend use of the
\lr cube where this problem does not occur at all.
We were unable to detect this problem in the ionized gas.  Since the spatial
changes in the gaseous nebula are smoother, this problem only affects point
sources but not the gas continuum or the emission lines.

\renewcommand{\topfraction}{1.0}
\setcounter{dbltopnumber}{6}
\renewcommand{\dbltopfraction}{1.0}

\section{Analysis of the ionized gas}\label{sec:warmgas}
\begin{figure*}
\begin{minipage}{0.48\linewidth}
\includegraphics[width=\linewidth]{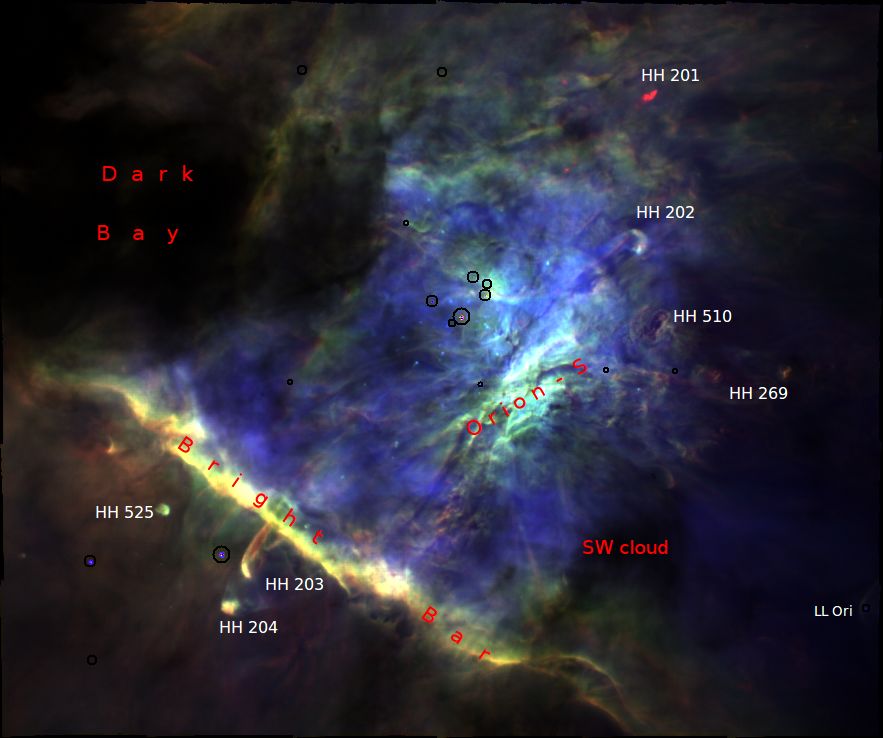}
\caption{Color composite using fluxes of three emission lines, with
         blue: \hb, green: \nii6584, and red: \sii6731.
         The brightest stars in the field are marked with black circles,
         large features are annotated in red text, other prominent features
         in white.}
\label{fig:colorlines}
\includegraphics[width=\linewidth]{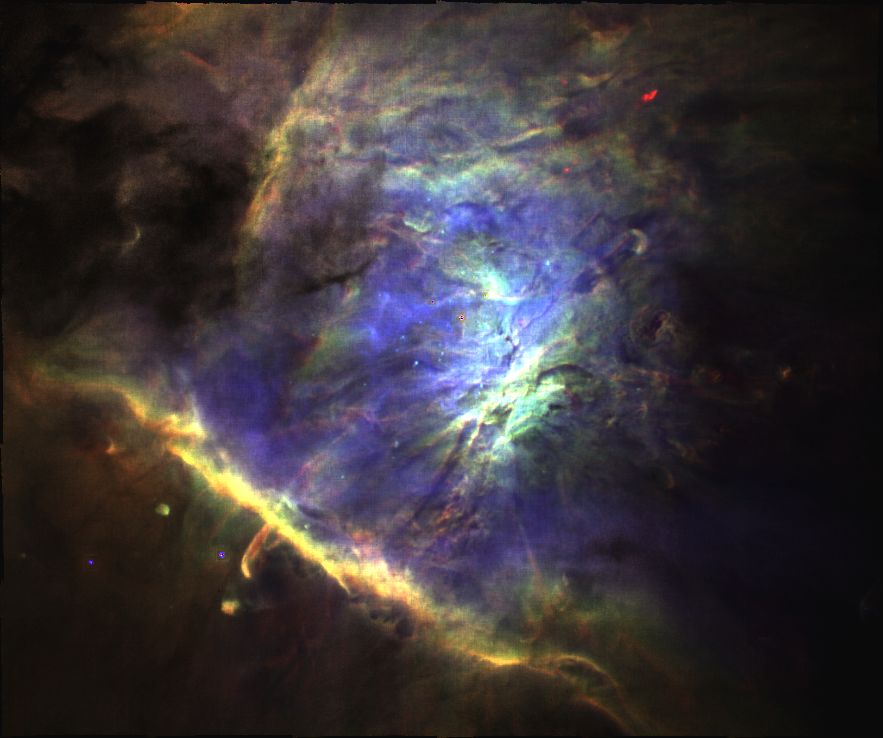}
\caption{Color composite, created from the same emission lines as
         Fig.~\ref{fig:colorlines} but using the reddening-corrected fluxes,
         using the extinction estimate from the Balmer decrement.}
\label{fig:colorderedd}
\end{minipage}
\hfill
\begin{minipage}{0.48\linewidth}
\includegraphics[width=\linewidth]{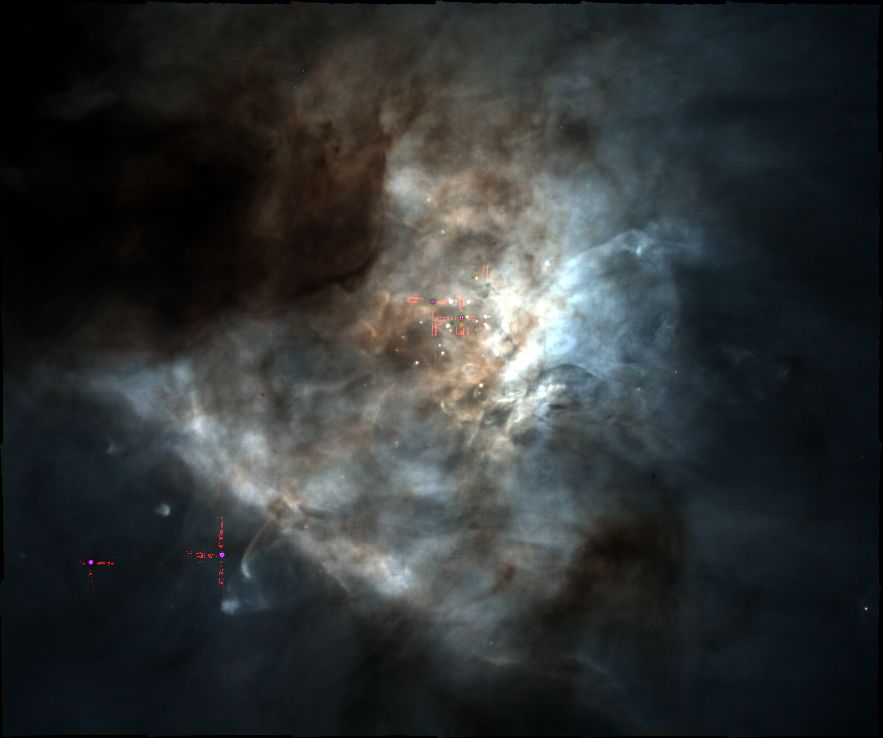}
\caption{Color composite, using emission line fluxes of red: Paschen 9, green:
         \ha, blue: \hb. In the red band, the effect of ghosts around bright
         stars is visible (cf.\ Fig.~\ref{fig:artifacts1}).}
\label{fig:colorhydrogen}
\includegraphics[width=\linewidth]{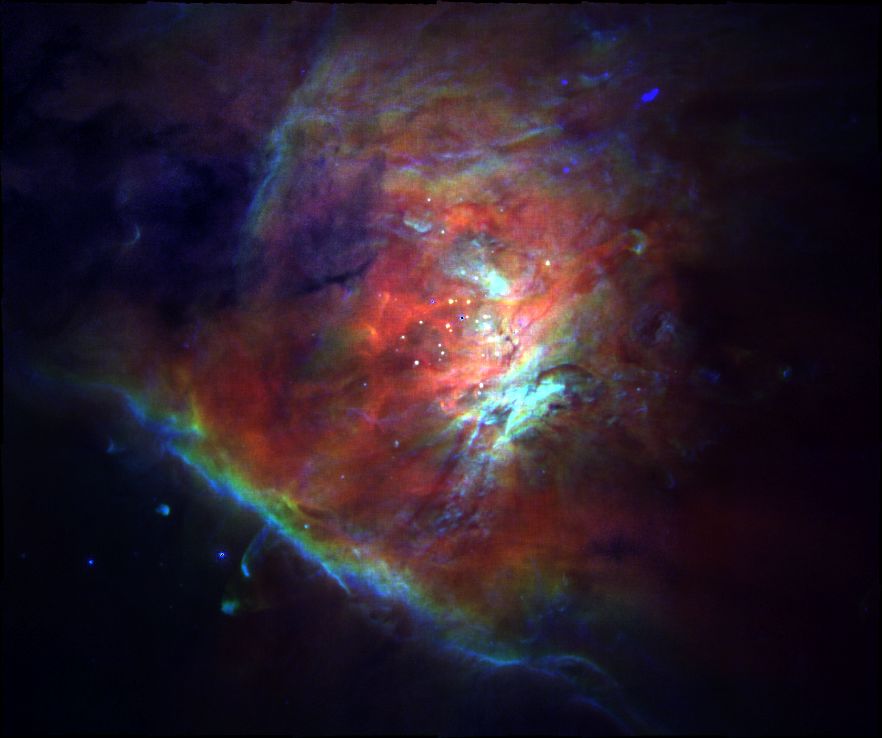}
\caption{False color image, using the three ionization levels of oxygen
         detectable with the MUSE data: red: \oiii 5007, green: \oii 7320,
         blue: \oi 6300, all corrected for extinction using the Balmer
         decrement.
         \oi is partially corrected for sky-line contribution.
         This representation shows the stratification of the nebula even more
         clearly than Fig.~\ref{fig:colorlines}.}
\label{fig:coloroxygen}
\end{minipage}
\end{figure*}

This section is devoted to show-case analysis methods that are possible with
the MUSE data. All maps we show here were created using Gaussian fits to single
emission lines. The fit was carried out on the \lr data, separately in each
spaxel (using the \hr data gives almost identical results). We allowed
variations in the flux of the line, its width, and the central position
(restricted to be within 5\,\AA\ of the expected zero-redshift center). The fit
also included a constant background offset.  Since we did not attempt to mask
stars before the fit, these show up as artifacts in some of the maps,
especially for Balmer lines.

At the spectral resolution of MUSE, the line \oi 6300 cannot be separated from
the sky line at the same wavelength. We therefore assumed that the nebular line
dominates the emission, but in the region of the Dark Bay (where the nebular
line is weak) and especially in MUSE exposures 31 to 36, the sky line was
stronger, leading to increased flux in the area covered by these exposures. We
modeled this as a linear multiplicative gradient, decreasing from 1.34 at the
left edge of the field to 1.0 at the approximate right edge of exposure 31+32 (cf.\
Fig.~\ref{fig:positions}), and divided it out. Since we ignored other lines
that coincide with telluric features, \oi 6300 is the only line that was
treated in this special way.

In Figures~\ref{fig:colorlines} to \ref{fig:coloroxygen} we show different
color representations, using the fluxes of three emission lines at a time,
extracted from the MUSE data.

In Fig.~\ref{fig:colorlines}, the selected lines trace different ionization
states in the main ionization front and hence different distances from
$\theta^1$\,Ori\,C \citep{2001ARA&A..39...99O}: \sii emission is produced in
the layer of the main ionization front on top of the molecular cloud, \nii at
intermediate distances, and \hb directly around the Trapezium cluster.  The
most striking features are the Bright Bar that runs across the image at the
bottom left, and the Orion-S region -- the brightest part of the nebula close
to the main ionizing source -- in the center.  Since the stars are visible only
as small and faint artifacts in this image, in the region to the SE of the
Bright Bar, only a few Herbig-Haro objects are prominently visible: the jets of
HH\,203 and HH\,204 \citep{1994ApJ...436..194O} next to each other and the more
roundish blob of HH\,524 \citep{2000AJ....119.2919B}.
In this image, we also note a spot with strongly enhanced \sii emission in the
upper right that appears as a red clump in Fig.~\ref{fig:colorlines}. This is
HH\,201, one of the ``bullets'' from the wide-angle Orion outflow
\citep{2003MNRAS.343..419G,2015arXiv150204711B}. Compared to the surrounding
material, it is similarly bright in \oi 6300. In the MUSE data, we detect a
secondary (blueshifted) component in the velocity field of these emission lines
in this region.

Fig.~\ref{fig:colorderedd} shows the same image but corrected for extinction
using the Balmer decrement (see Sect.~\ref{sec:exti}).
Fig.~\ref{fig:colorhydrogen} combines the emission line fluxes of three
hydrogen lines spread over almost the full MUSE wavelength range: Paschen9 (=
Pa$\zeta$) at 9229.7\,\AA, \ha 6562.8\,\AA, and \hb 4861.3\,\AA. This is the
image before correcting for extinction, the reddening-corrected version (not
shown) is devoid of color.

Fig.~\ref{fig:coloroxygen} shows three oxygen ions accessible using the MUSE
data: \oiii 5007 represents the highest ionization state, the hottest region of
the nebula and is colored red, \oii 7320 appears for the intermediate
ionization state, and \oi 6300 for the coldest gas. For this image we used the
extinction-corrected fluxes measured for the emission lines. The different
extent of the three states and the diversity in structure visible are both due
to the stratification of the nebula. This is especially visible at the Bright
Bar.

\begin{figure*}
\centering
\includegraphics[width=\linewidth]{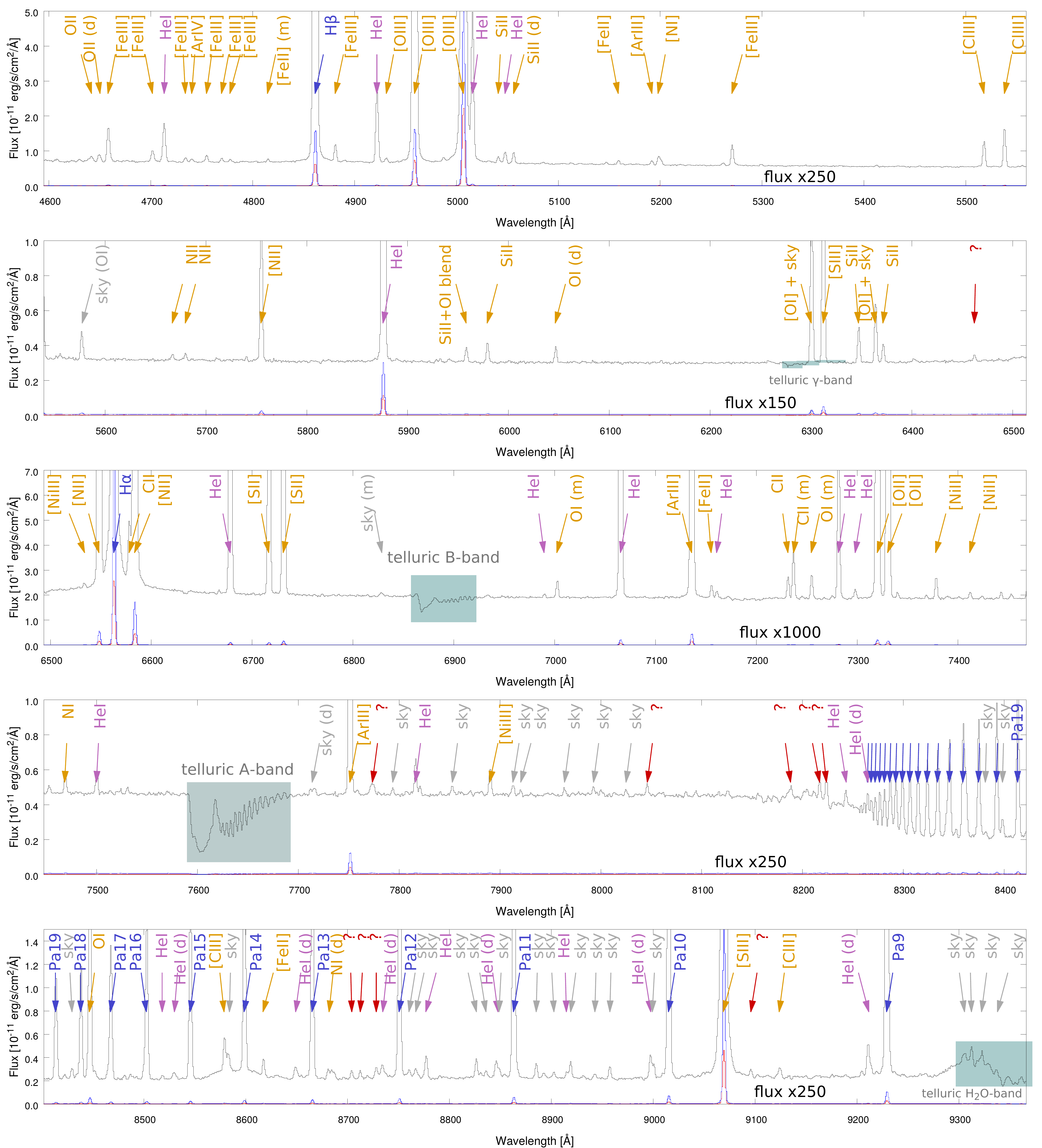}
\caption{Full spectrum of the MUSE dataset, extracted at the approximate
         positions of slit1 of \citet{BFM+91} (red) and at the optical
         slit of \citet{1992ApJ...389..305O} (blue). The ``slit1''-spectrum is
         shown again in black, multiplied with the factor at the bottom of each
         panel, to highlight fainter lines.
         Note that the different panels have different absolute scales.
         Line identifications from the same two references are shown, hydrogen
         lines are in blue (lines Pa20 to Pa35 are marked but not annotated),
         helium lines in violet, metal lines in orange, and lines where the sky
         contribution dominates are grey. Unresolved line doublets and
         multiplets are marked with '(d)' and '(m)', respectively.
         A few faint lines without identification are marked in red.}
\label{fig:fullspectrum}
\end{figure*}

Fig.~\ref{fig:fullspectrum} presents two nebular spectra over the full
wavelength range extracted from the MUSE data.  At the blue end of the
spectrum, the first visible emission line is \ion{O}{ii} 4642, but in the
cube, several other, even fainter lines are detected. Hydrogen Paschen lines
can be discriminated from Pa9 to Pa35 at the red end.

Like the extraction regions from the cube, the line identification in
Fig.~\ref{fig:fullspectrum} were taken from \citet{BFM+91} and
\citet{1992ApJ...389..305O}.  We can identify lines down to a flux of about
$0.005 \times F_\mathrm{\hei 6678}$ listed in their reference tables. However,
a few lines visible in our spectrum are not listed in these reference sources
and are not known sky emission lines. They remain unidentified and are
specially marked in our figure.

As a first demonstration of what is possible with this new dataset, we analyze
the nebular emission in a spatially resolved (pixel by pixel) manner.

\subsection{Extinction}\label{sec:exti}
\begin{figure}
\centering
\includegraphics[width=\linewidth]{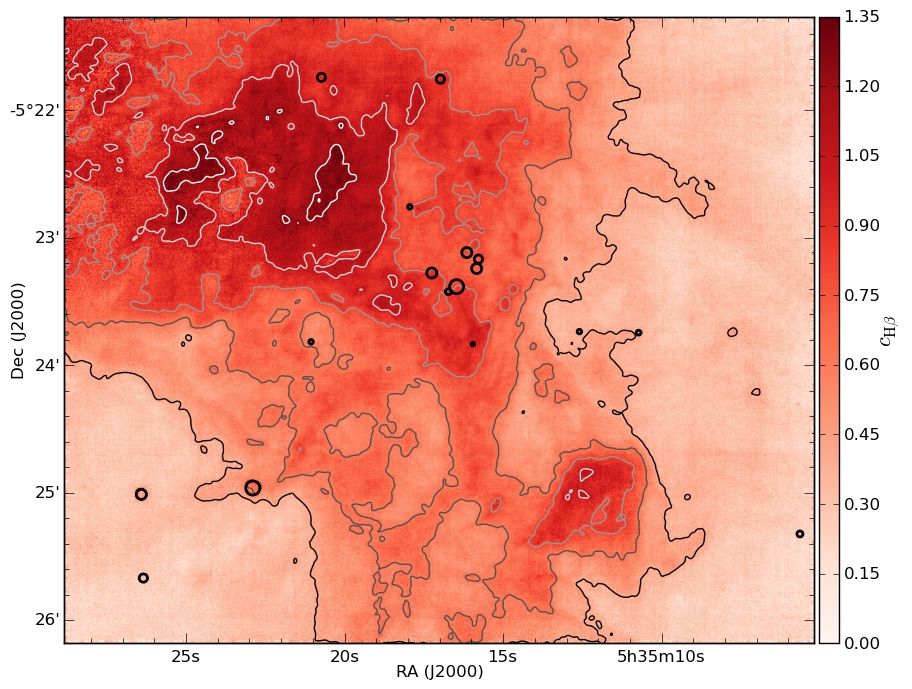}
\caption{Extinction map of the central Orion Nebula, as derived from the
         \ha to \hb flux ratio, displayed as $\chb$.
         Bright stars are marked again with black circles, the smoothed contours
         are $0.4 \le \chb \le 1.2$ in steps of 0.2.}
\label{fig:extinct}
\end{figure}

We follow \citet{ODH10} by selecting an intrinsic H$\alpha$ to \hb flux
ratio of 2.89, for case B and $T_e=9000$\,K
\citep[see][]{1995MNRAS.272...41S,OF05}.
We interpolate the flux measurement of the Balmer lines at the positions of
bright stars (where the Balmer decrement could not be measured reliably), and
then input them as well as the reference value into the {\sc PyNeb} software
\citep[v1.0.9,][]{2013ascl.soft04021L,PyNeb}, and derive a reddening map using
the extinction curve of \citet{CCM89} as refined for the Orion Nebula by
\citet{2007ApJ...655..299B} with $R_V=5.5$. This map is displayed in
Fig.~\ref{fig:extinct} as $\chb$. The contour levels displayed there correspond
to $0.9 \lesssim A_V\lesssim 2.7$\,mag or -- when applying the relation of
\citet{1978ApJ...224..132B} -- to column densities of $9.6\times10^{20} \lesssim
N(\mathrm{HI}+\mathrm{H}_2) \lesssim 2.9\times10^{21}\,\mathrm{cm}^{-2}$.  As
expected, the regions of the ``Dark Bay'' and the SW cloud
\citep{2009AJ....137..367O} show the strongest extinction while e.\,g.\ the
region south-east of the ``Bright Bar'' as well as the western arc-minute of
our field exhibit very moderate reddening.
By contrast with (sub)mil\-li\-meter emission maps
\citep[e.\,g.][]{1999ApJ...510L..49J}, the foreground dust structure appears
smooth and devoid of significant substructure.  Such substructure is most
commonly seen in self-gravitating gas clouds that form filamentary and clumpy
structures on scales from the parent cloud down to individual protostars
\citep[e.\,g.][]{2006ApJ...653..383J,2013ApJ...763...57T}.  The lack of this
clumping in our extinction map suggests that this foreground material is not
self-gravitating.

While qualitatively similar to the reddening map derived by
\citet{2000AJ....120..382O} from a radio-to-optical surface brightness comparison
we note that we only reach peak values of $\chb=1.35$ in the Dark
Bay, whereas O'Dell \& Yusef-Zadeh found values up to 2.0. The comparison to their
Balmer decrement-derived map, however, shows very similar absolute trends over
the smaller area covered by them, such that in the vicinity of the Trapezium
$\chb$ reaches 0.8 but is only about 0.5 between the SW cloud and
the Trapezium.

We use this map to correct all measured emission line fluxes for reddening.  We
then also use the extinction-corrected flux maps to recreate a "dust-free"
version of Fig.~\ref{fig:colorlines} in Fig.~\ref{fig:colorderedd} as well as
Fig.~\ref{fig:coloroxygen} (for which no uncorrected counterpart is shown). It
is apparent from the latter image that the reddening correction based on the
Balmer-decrement is imperfect, since the Dark Bay and the SW cloud get more
transparent but do not disappear.

\subsection{Emission line maps}
\begin{figure*}
\begin{minipage}{0.48\linewidth}
\includegraphics[width=\linewidth]{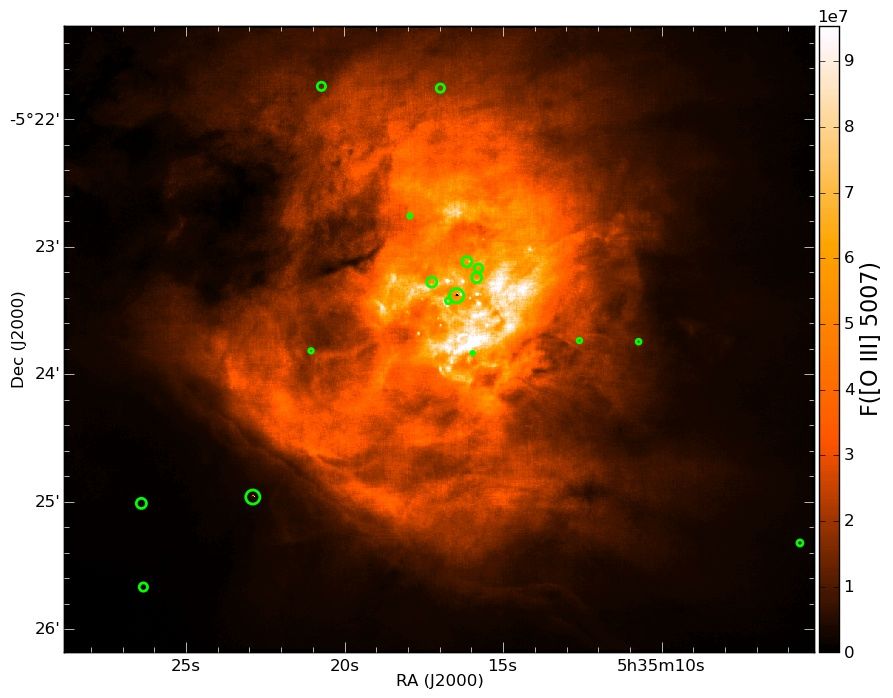}
\caption{Reddening-corrected flux map of \oiii 5007.}
\label{fig:emlinemap1}
\includegraphics[width=\linewidth]{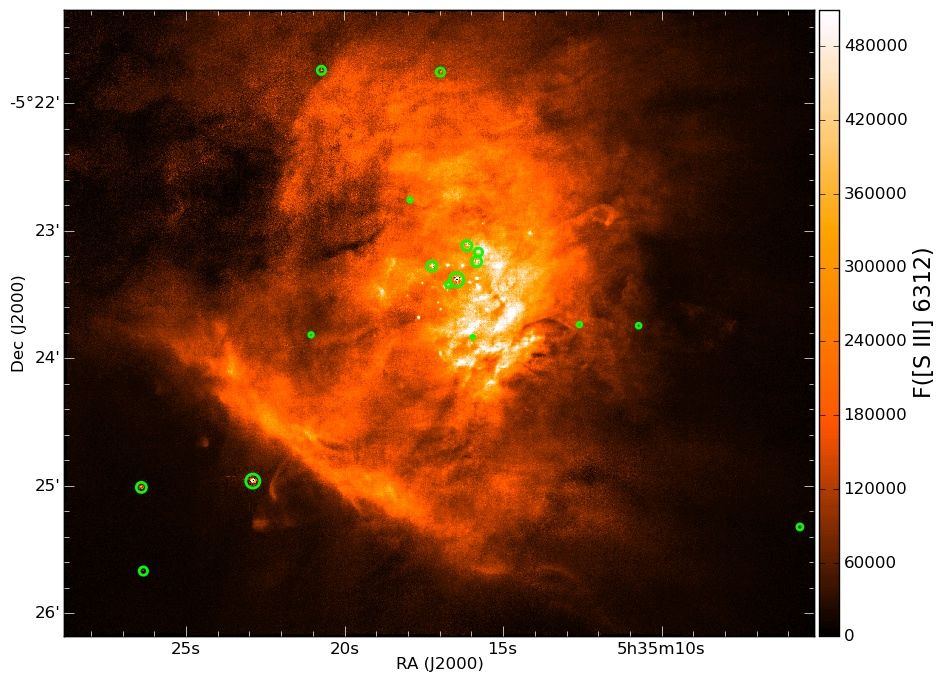}
\caption{Reddening-corrected flux map of \siii 6312.}
\label{fig:emlinemap2}
\end{minipage}
\hfill
\begin{minipage}{0.48\linewidth}
\includegraphics[width=\linewidth]{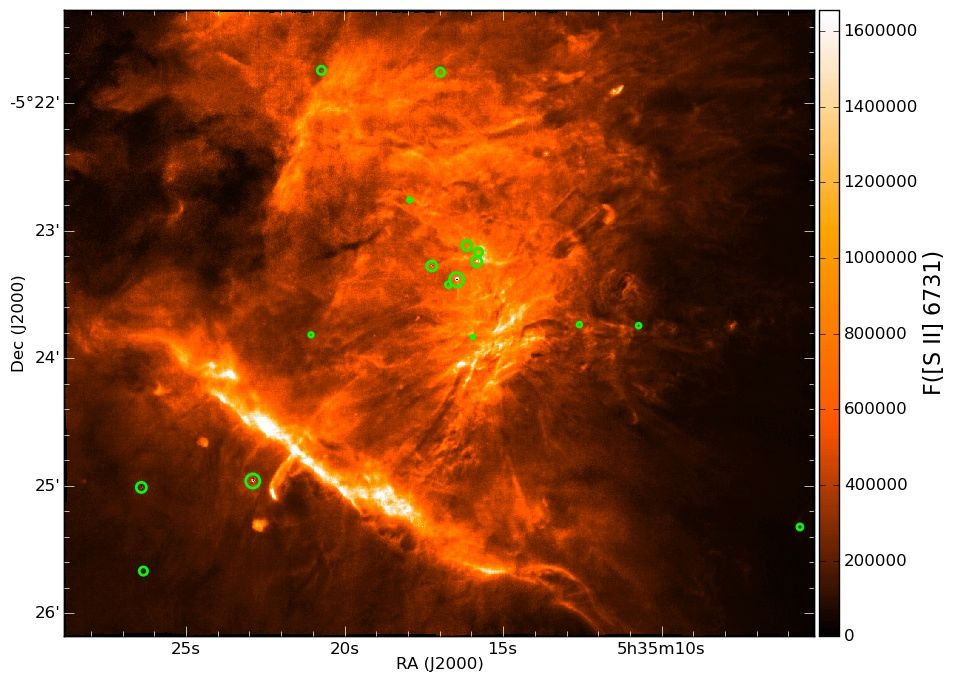}
\caption{Reddening-corrected flux map of \sii 6731.}
\label{fig:emlinemap3}
\includegraphics[width=\linewidth]{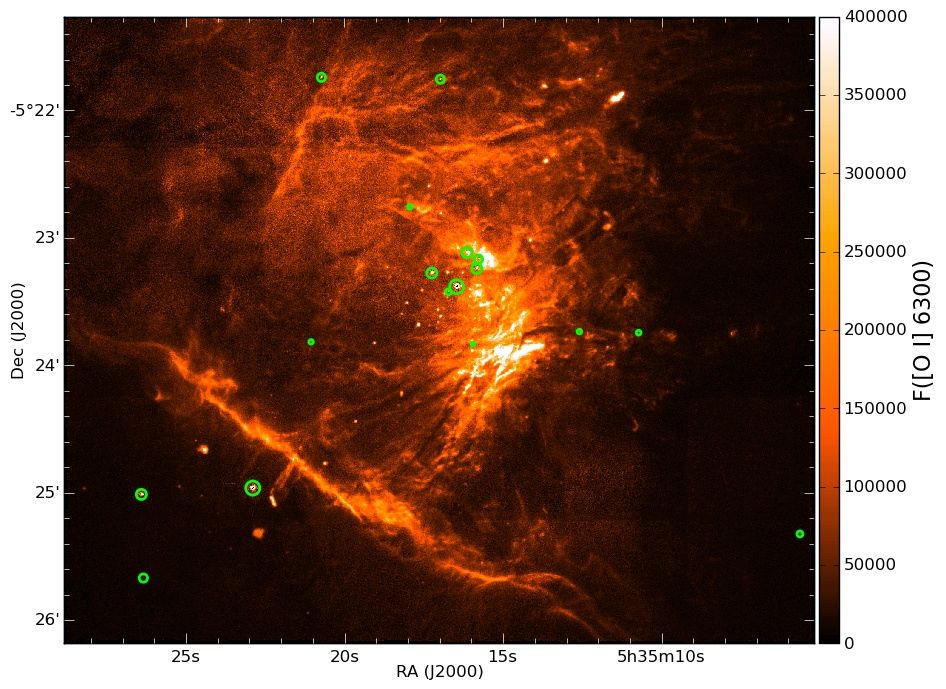}
\caption{Reddening-corrected flux map of \oi 6300, partially corrected for
         sky-line contribution.}
\label{fig:emlinemap4}
\end{minipage}
\end{figure*}

Figures \ref{fig:emlinemap1} to \ref{fig:emlinemap4} show the Orion Nebula in
dedicated emission lines as representatives of different ionization stages.
The images of the Orion Nebula are more compact and diffuse in the higher
ionized lines \oiii 5007 and \siii 6312 than in the lower ionized lines \sii
6731 and \oi 6300.

The fact that the images in the low ionization lines such as \oi are much more
structured than in the higher ionization lines is caused by the fact that the
low ionization lines only form over a very narrow range of physical conditions
in the nebula in comparison to the high ionization lines.
Such effects are known e.\,g.\ from images of planetary nebulae taken in the
light of different emission lines \citep{OF05}.
Neutral gas is also often intrinsically more structured as the lower temperature,
and therefore pressure, means that the neutral gas is more easily compressed.

\subsection{Diagnostic maps}
With the amount of emission lines present in the MUSE data and the reddening
correction derived in Sect.~\ref{sec:exti}, we can easily create flux ratio
images, i.\,e.\ diagnostic maps.

\begin{figure*}
\begin{minipage}{0.48\linewidth}
\includegraphics[width=\linewidth]{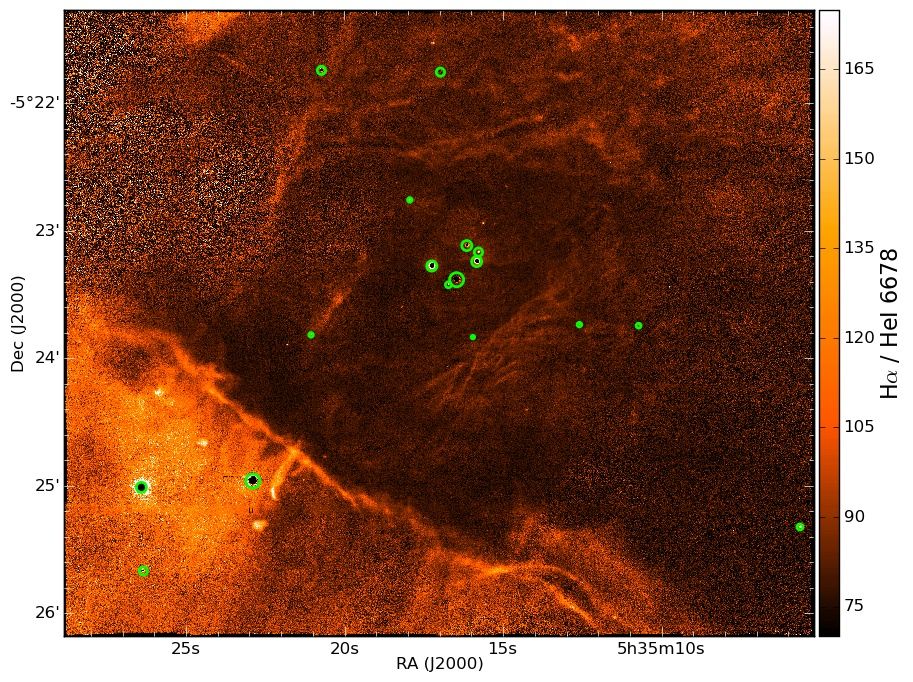}
\caption{Reddening-corrected emission line ratio map of \ha/\hei 6678.}
\label{fig:diag1}
\includegraphics[width=\linewidth]{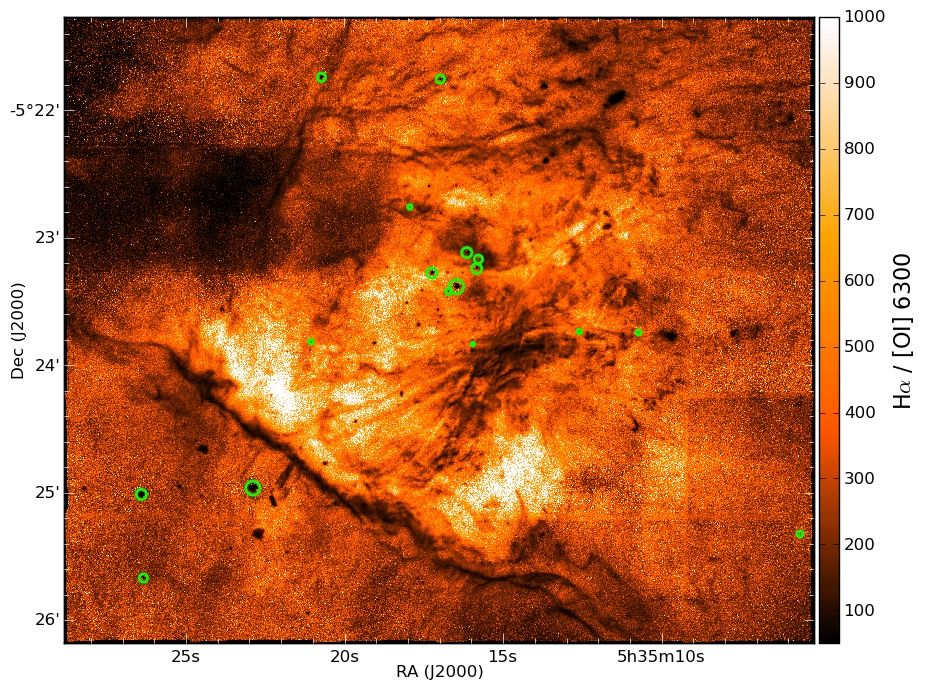}
\caption{Reddening-corrected emission line ratio map of \ha/\oi 6300, partially
         corrected for sky-line contribution to \oi 6300.}
\label{fig:diag2}
\end{minipage}
\hfill
\begin{minipage}{0.48\linewidth}
\includegraphics[width=\linewidth]{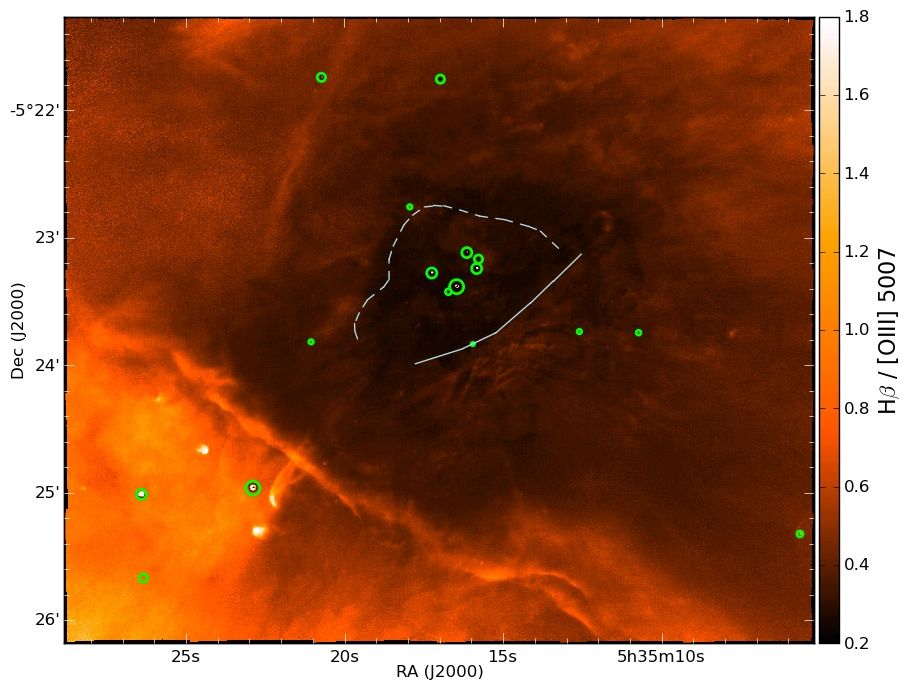}
\caption{Reddening-corrected emission line ratio map of \hb/\oiii 5007.
         The \oiii shell and the SE-NW ionization boundary discussed by
         \citet{2009AJ....137..367O} are marked.}
\label{fig:diag3}
\includegraphics[width=\linewidth]{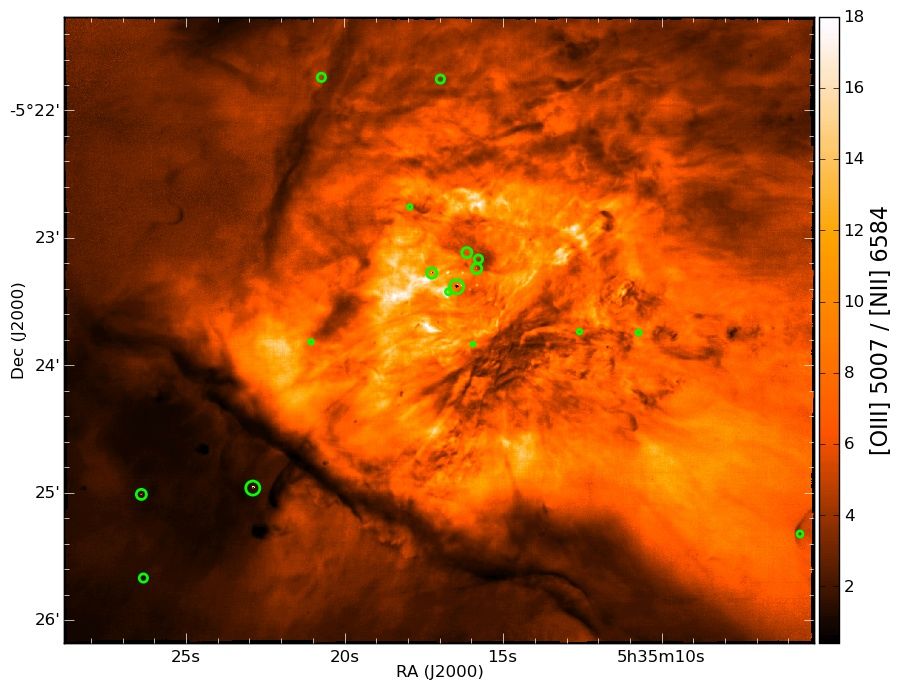}
\caption{Reddening-corrected emission line ratio map of \oiii 5007/\nii 6584.}
\label{fig:diag4}
\end{minipage}
\end{figure*}

The diagnostic images (Fig.~\ref{fig:diag1} to \ref{fig:diag4}) -- based on
different line intensity ratios -- highlight different regions in the Orion
nebula.
The figures based on the \ha/\hei 6678 (Fig.~\ref{fig:diag1}) and \hb/\oiii
5007 (Fig.~\ref{fig:diag3}) line ratios highlight the innermost and most
strongly ionized structures of the nebula surrounding the central Trapezium
stars. These line ratios are mainly indicators of the mean level of ionization
and temperature \citep{OF05}.
The discovery of a central \oiii shell surrounding the Trapezium stars has been
reported by \citet{2009AJ....137..367O}, indicating a stationary
high-ionization structure. We also see hints of this structure in our \hb/\oiii
5007 image (Fig.~\ref{fig:diag3}, cf.\ Fig.~3 of O'Dell et al.\ 2009) and may
even detect this \ha/\hei 6678 map (Fig.~\ref{fig:diag1}). Furthermore,
extended shell structures are clearly visible towards the north.

In the \oiii 5007/\nii 6584 image (Fig.~\ref{fig:diag4}) the darker Orion-S
region stands out south-west of the Trapezium. This foreground Orion-S complex
hosts embedded stars that are sources of many large-scale optical outflows
\citep{2009AJ....137..367O}. The bright elongated region towards the outer
south-west highlights the shocked wind region. Here the gas flows towards the
low-density end in a so-called champagne flow \citep{2006ApJS..165..283A}.  In
addition, the bow shock connected to the T-Tauri star \object{LL\,Ori} (see spectrum in
Fig.~\ref{fig:stellarspectra}) sticks out to the outermost southwest.
Furthermore, towards the west a loop structure pops up. Similar structures can
be recognized as well in highly processed 20\,cm-continuum images taken with
the VLA \citep{1990ApJ...361L..19Y}.

The \ha/\oi 6300 image (Fig.~\ref{fig:diag2}) shows sharp extended structures
in the outer (cooler) regions of the Orion Nebula.

\subsection{Electron temperature / density}
We again use {\sc PyNeb} to cross-iterate electron density ($\Ne$) and
electron temparature ($\Te$) of the ionized gas, using
extinction-corrected line ratios. We use the temperature sensitive line ratios
\nii 5755/6548, \nii 5755/6584, and \siii 6312/9069
together with the density sensitive ratios \sii 6731/6716 and
\cliii 5538/5518, and let the iteration start at
$\Te=10000$\,K.
The assumptions for using PyNeb are, that the $n$-level approximation that this
tool is based on \citep[also see][for a 5-level precursor]{1987JRASC..81..195D}
can describe the ionized states of the gas involved, and that the gas along the
line of sight is sufficiently homogeneous for the emission lines to represent
the luninosity-weighted physical state of the ionized gas in each spatial
element.

We derive maps of $\Te$ from \nii (averaged from 5755/6548
and 5755/6584, cross-iterated with \sii) and \siii 6312/9069
(cross-iterated with \sii\footnote{The map cross-iterated against
  \cliii gives very similar temperatures, but has much lower quality,
  due to many more pixels with non-converging iterations.}),
and $\Ne$ from \sii 6731/6716 and \cliii 5538/5518.

Since we did not attempt to disentangle stellar continuum and gas or mask
positions of stars, many strong small-scale features in these maps
may be artifacts. This can be easily verified using a continuum part of the
spectral range. The brightest stars are therefore marked on the corresponding
maps.

\begin{figure*}
\begin{minipage}{0.48\linewidth}
\includegraphics[width=\linewidth]{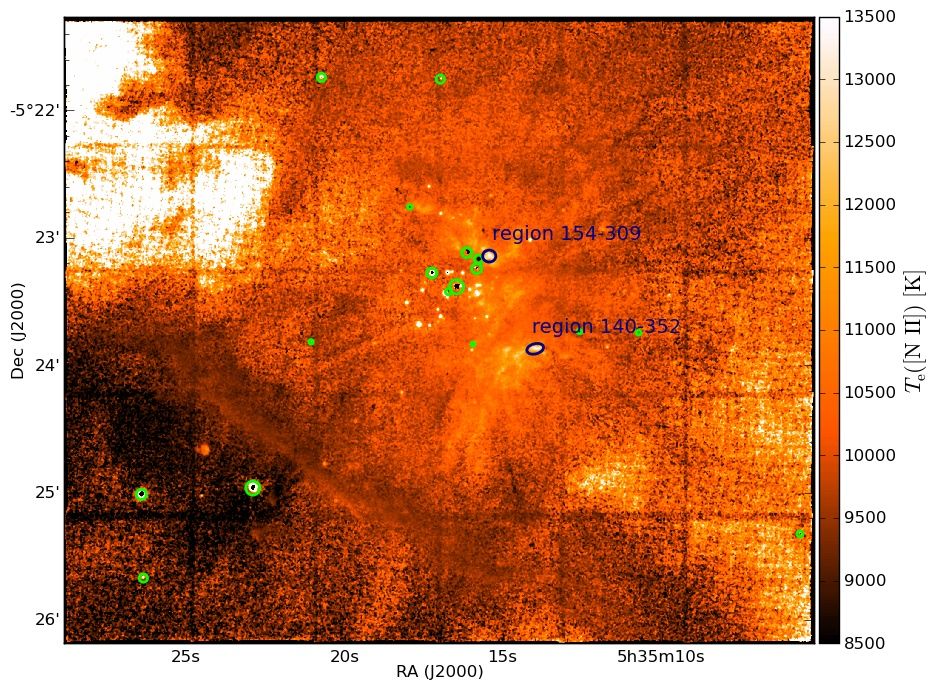}
\caption{\nii-derived $\Te$-map of the central Orion Nebula, smoothed by a
         median filter of 5$\times$5 pixels box width, displayed in linear
         scaling. See text for more details.}
\label{fig:Te_NII}
\includegraphics[width=\linewidth]{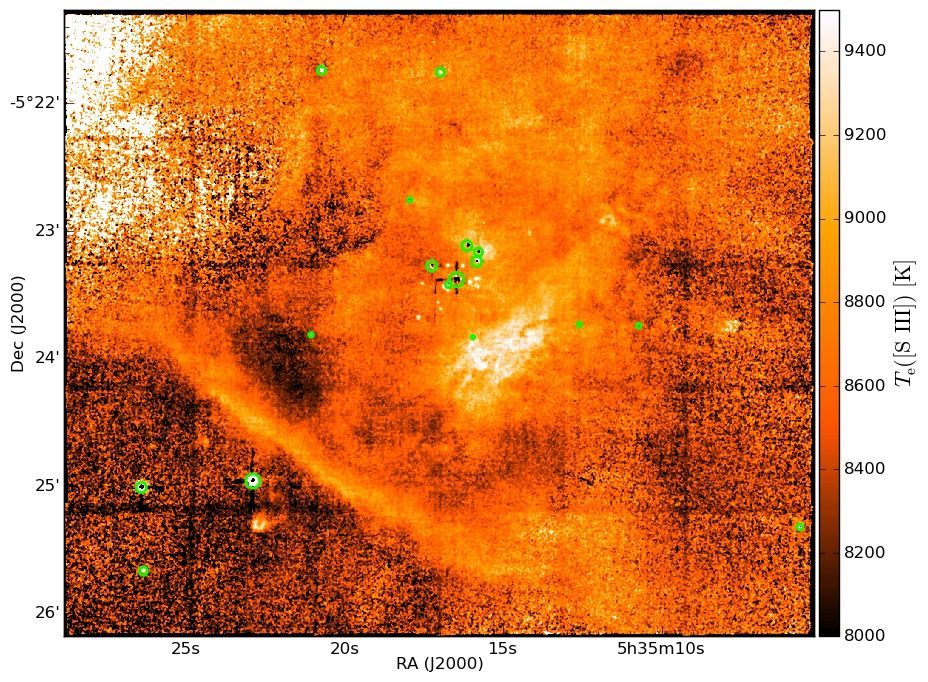}
\caption{\siii-derived $\Te$-map of the central Orion Nebula, smoothed by a
         median filter of 5$\times$5 pixels box width, displayed in linear
         scaling. See text for more details.}
\label{fig:Te_SIII}
\end{minipage}
\hfill
\begin{minipage}{0.48\linewidth}
\includegraphics[width=\linewidth]{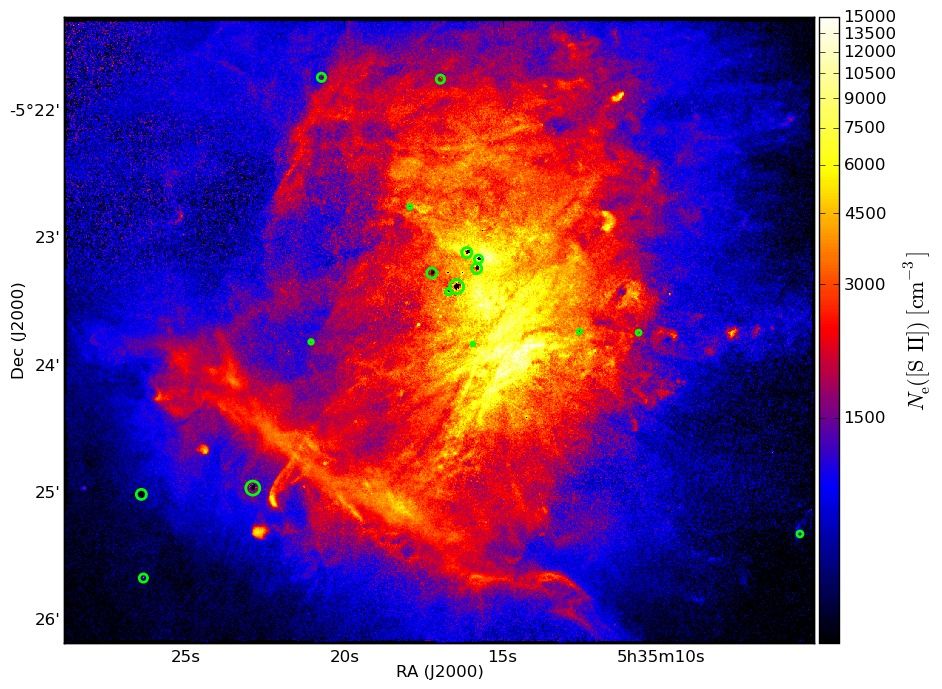}
\caption{\sii-derived $\Ne$-map of the central Orion Nebula, smoothed by a
         median filter of 3$\times$3 pixels box width, displayed in asinh
         scaling. See text for more details.}
\label{fig:ne_SII}
\includegraphics[width=\linewidth]{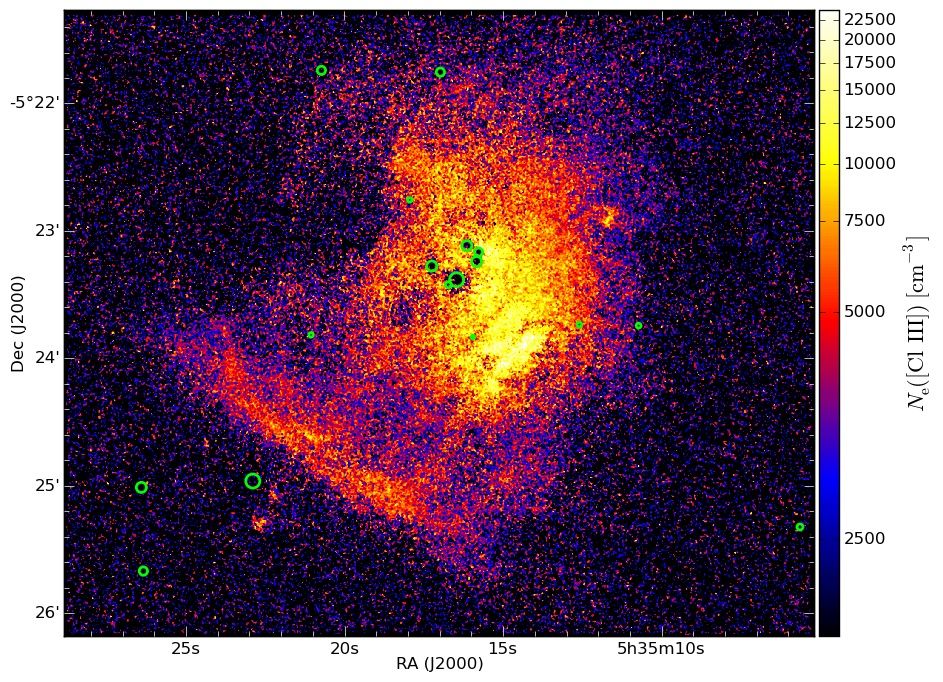}
\caption{\cliii-derived $\Ne$-map of the central Orion Nebula, smoothed by a
         median filter of 5$\times$5 pixels box width, displayed in asinh
         scaling. See text for more details.}
\label{fig:ne_ClIII}
\end{minipage}
\end{figure*}

In the \nii-derived $\Te$-map in Fig.~\ref{fig:Te_NII}, the
hottest regions are 154-309 \citep[in the coordinate system
of][]{1994ApJ...436..194O},
about 21\arcsec\ NW from $\theta^1$\,Ori\,C (marked in Fig.~\ref{fig:Te_NII}),
and 140-352,
47\arcsec\ SW of $\theta^1$\,Ori\,C, both reaching $\sim$13000\,K. The coldest
region appears to be beyond the Bright Bar, around $\theta^2$\,Ori\,A and
$\theta^2$\,Ori\,B, with $\Te \sim 8500$\,K.  That $\Te$ is
found to be high in the Dark Bay may be a problem of imperfect extinction
correction.
A grid-like pattern is visible in this map, caused by the $\Te$ estimate to be
about 200-300\,K lower in the regions where multiple exposures overlap with
adjacent pointings. This is caused by a slight systematic bias of the weaker
line \nii 5755 and can be viewed as a represention the systematic uncertainty
of these maps.

The \siii-derived $\Te$-map in Fig.~\ref{fig:Te_SIII} shows a
different behavior in comparison to the \nii-derived map.
The hottest regions are some of the Herbig-Haro shocks, e.\,g.\ HH\,204
reaching $\sim$9200\,K, and the Bright Bar and the region SW \citep[called the
``Orion-S'' feature by][]{2009AJ....137..367O} of the Trapezium are hotter than
the surrounding nebula.  The coldest regions in this ionization layer with
about 7800\,K are between the Bright Bar and the Dark Bay and between the
Trapezium and the Dark Bay.

\siii was previously used by \citet{1974PASP...86..211F} to derive $\Te =
9700\pm1000$\,K in a slit of $150\arcsec\times15\arcsec$ in an unspecified
location. Assuming that they pointed within the brightest part of the nebula
but not on the Trapezium stars, their result agrees with ours.
Again, the extremely high $\Te$ values derived in the Dark Bay may be an
artifact. Similar to $\Te(\nii)$, the grid-like structure originates in the
flux measurement of the fainter line \siii 6312.

Due to the density of known sources in the Orion Nebula and changes of the
absolute world coordinate systems used in the literature, it is not always
clear which known objects are related to features in our data. However, at
least a few of the compact high-$\Te$ peaks visible in Fig.~\ref{fig:Te_NII}
and \ref{fig:Te_SIII} around the location of the Trapezium cluster can be
identified with known young stellar objects and proplyds, e.\,g.\ 170-337
\citep{1994ApJ...436..194O}, d141-301 and j177-341 \citep{2000AJ....119.2919B}.

The electron density as derived using the \sii doublet (see
Fig.~\ref{fig:ne_SII}) is consistent with the map derived by
\citet{1992ApJ...399..147P}, but shows features with higher spatial resolution.
$\Ne$ varies between $\sim500\,\mathrm{cm}^{-3}$ at the edge of the field
and densities in excess of $10\,000\,\mathrm{cm}^{-3}$ in the Orion-S region.

The layer of the ionization front showing \cliii emission has an even higher
density, reaching up to $\Ne\approx25\,000\,\mathrm{cm}^{-3}$ in parts of the
Orion-S region, as shown on Fig.~\ref{fig:ne_ClIII}. The lowest values derived
from the \cliii doublet give $\Ne\approx4000\,\mathrm{cm}^{-3}$, just north
of the Bright Bar.
That $\Ne$ estimated from \cliii gives higher densities than derived from \sii
was qualitatively already presented by \citet{2013hsa7.conf..594N}. However,
their scale does not allow a direct comparison to our map.
For most of the field of view, the densities from \cliii are too noisy to
distinguish more than genereral trends with position.

The range in derived electron temperatures and densities prompted
\citet{SCV+07} to compute the extinction values at each position using the
physical gas conditions inferred by them. However, the dependency of the Case B
Balmer decrement on the densities is weak (for the values derived here, it only
changes by 1\%). Since the temperature estimate is not independent of the
reddening correction, we would need to add another layer of iterations. The
expected changes are small (changes of the derived Balmer decrement reference
value are at most 3\%), we therefore prefer to not carry the analysis beyond
this point.

\subsection{Velocity field}
\begin{figure}
\centering
\includegraphics[width=\linewidth]{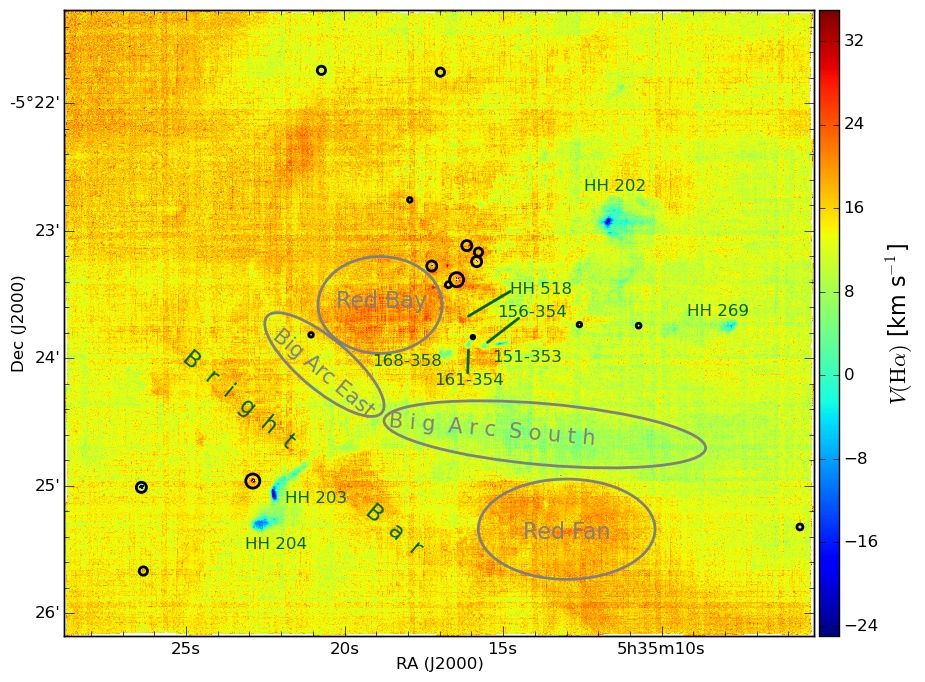}
\includegraphics[width=\linewidth]{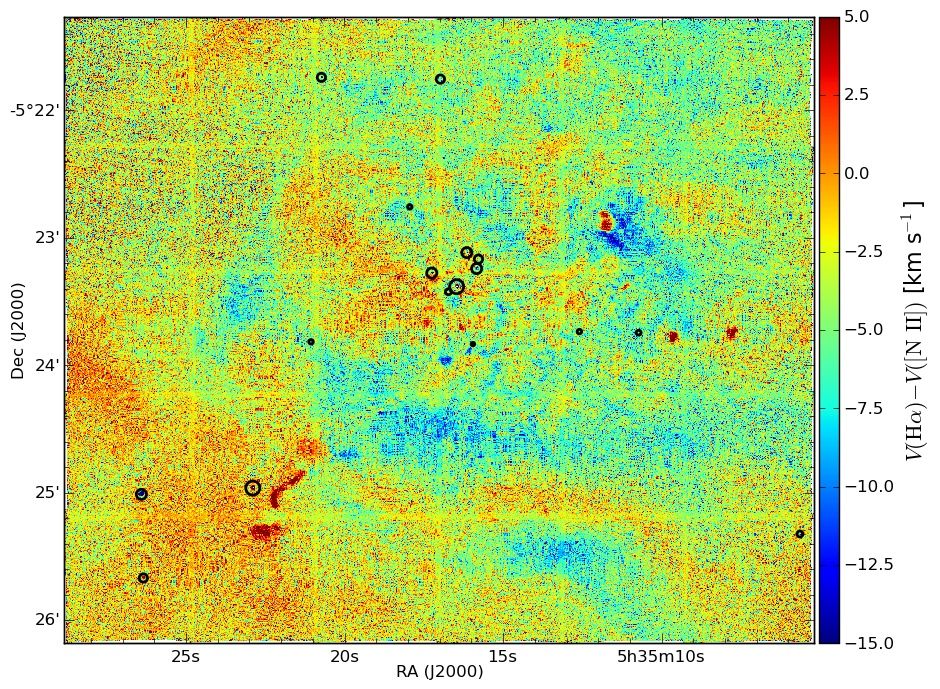}
\includegraphics[width=\linewidth]{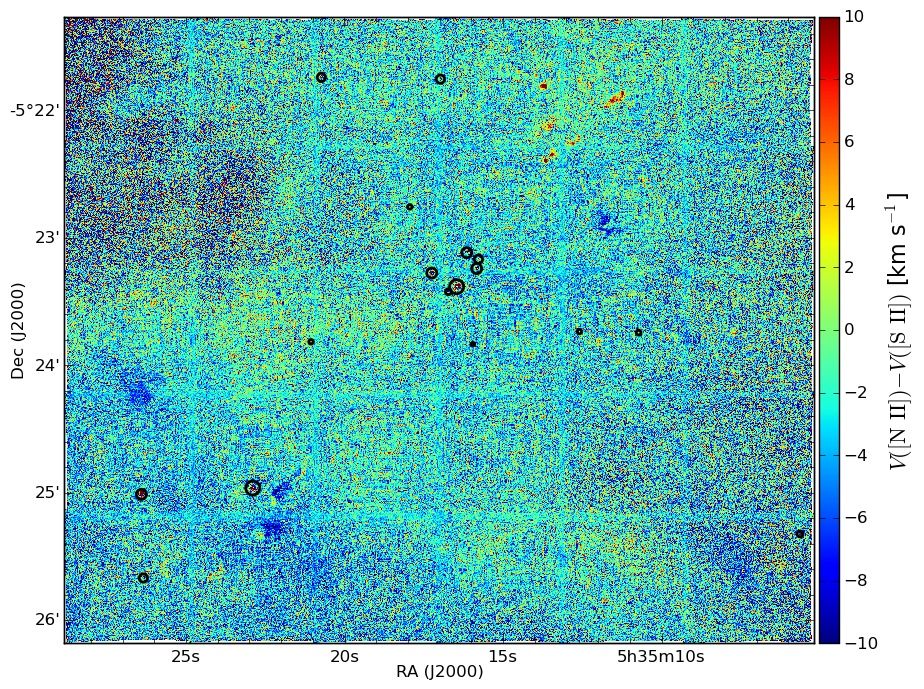}
\caption{Velocity field as traced by the lines \ha ({\bf top}, with respect to
         barycentric velocity), \nii 6584 relative to \ha ({\bf middle}), and
         \sii 6731 relative to \nii 6584 ({\bf bottom}).
         Note that the linear color scales are different for all these plots
         and that the range of the middle plot is asymmetric. The locations of
         the brighter stars in the field are marked as black circles, further
         noticeable features are annotated or marked in green or gray on the
         \ha velocity map.}
\label{fig:velos_full}
\end{figure}

Fig.~\ref{fig:velos_full} shows the velocity field recovered from Gaussian
centroids of the emission lines \ha (absolute velocities against barycentric
zeropoint), \nii 6584 (relative to \ha), and \sii 6731 (relative to \nii) for
the full field mapped by MUSE.

The strongest features in velocity space for \ha are the Herbig-Haro jets,
especially the blueshifted objects HH\,202, HH\,269, HH\,203, and HH\,204
(marked on the \ha velocity image in Fig.~\ref{fig:velos_full}, top panel). A
few more small-scale features are visible just south of the Trapezium stars,
using the coordinate-based designation system for M\,42 as invented by
\citet{1994ApJ...436..194O}.  The Bright Bar is visible as region of slightly
enhanced velocity; more regions of visibly different redshift are marked with
annotated grey ellipses on the H$\alpha$ velocity map of
Fig.~\ref{fig:velos_full}.
The two higher velocity ($17 \lesssim v_\mathrm{H\alpha} \lesssim 21$\kms)
regions were discussed by \citet{2007AJ....133..952G}, they were named "Red
Bay" for the region east of the Trapezium, and "Red Fan" for the part just
north of the western end of the Bright Bar. The elongated structure of lower
velocities between Red Bay, Bright Bar, and Red Fan was discussed before in
\citet{2004AJ....127.3456D} and named "Big Arc" with a less
pronounced eastern component ($11 \lesssim v_\mathrm{H\alpha} \lesssim 14$\kms
in our data) and a stronger and longer southern part ($3 \lesssim
v_\mathrm{H\alpha} \lesssim 14$\kms in the MUSE data).
This map can be compared to the \ha velocity map shown by
\citet[][their Fig.~13]{2008RMxAA..44..181G}. In both datasets, the large scale
features show up in a very similar way. Even smaller features, like the
velocity dips south of the Trapezium (especially at 168-358 and 161-354) are
visibly similar. The absolute velocity values are slightly different between
our data and theirs, with their value $\sim2.5$\kms higher. This difference is
within the combined error estimates of both datasets.\footnote{We computed our
  velocity map with respect to $\lambda_\mathrm{H\alpha} = 6562.791$\,\AA, see
  \citet{1999A&AS..135..359C}.}

To highlight differences in velocity derived from our data for different ions,
we also show the velocities derived for \nii 6584, subtracted from the
\ha velocity field (Fig.~\ref{fig:velos_full}, middle panel). The velocity
offset shown in this map is around $-3.5\pm4.8$\,\kms.
Prominent features in this map are HH\,202 to the west of the Trapezium and
HH\,203 and HH\,204 beyond the Bright Bar, but on careful investigation it
shows a plethora of other features, like blueshifted compact features not
related to prominent Herbig-Haro objects, fainter filamentary structures around
the Orion-S region and smoother large-scale changes. The Big Arc again shows
up prominently, as blue region.
Constructing the same velocity difference map from the data of
\citet{2008RMxAA..44..181G} shows a very close match (W.~Henney, priv.~comm.).
Their average velocity difference is $-3.7$\kms (their Table 2), very close to
the value computed here, even though the area covered by their data is not
identical.

The bottom panel of Fig.~\ref{fig:velos_full} shows the velocity difference map
of \sii 6731 compared to \nii 6584. The velocity offset between these lines is
much smaller, $-2.5\pm4.8$\,\kms.
Large-scale features are less pronounced than in the previous velocity maps,
and when interpreting them one should keep in mind that the scale shown is far
below the velocity resolution achievable with MUSE.  Nevertheless, the velocity
differences beyond the Bright Bar, i.\,e.\ in the region around HH\,203 and
204, seem systematically negative. We measure mean values of about $-4\,$\kms
in that part of the nebula, so that the velocity of the \sii line is higher in
that region than \nii, in contrast to the central part where both lines appear
to have similar velocities. Another possible velocity difference occurs within
the Dark Bay, but the noise in that region is too large to be certain.\footnote{Both
  in this bottom panel and in the middle panel discussed above, the grid-like
  structure of the original pointings is visible.  This is due to higher S/N in
  the small regions of overlap between. When smoothed spatially, these
  structures disappear. The velocities measured within and just outside these
  overlap regions agree well within the error bars.}
This map can again be compared to the measurements of
\citet{2008RMxAA..44..181G}. Although their area is slightly smaller, they also
find a comparable mean velocity difference of $-0.7$\kms. In the velocity
difference map created from their data, one can also determine values of zero
in the Red Bay and the Red Fan, like in the MUSE data, but a slightly smaller
difference beyond the Bright Bar ($-1.3$\kms, W.~Henney, priv.~comm.).
Small-scale features are again the well-known jets of HH\,203 and HH\,204 as
well as HH\,202. But here, some of the outflows from Orion-BN/KL as recently
observed in the near-infrared by \citet{2015arXiv150204711B} show up
prominently as red patches to the north-west of the Trapezium, especially
HH\,201, HH\,209, and features near HH\,208.

\section{Conclusions}\label{sec:concl}
We successfully demonstrated the capabilities of the integral-field
spectrograph MUSE instrument with a new dataset representing imaging
spectroscopy of the Huygens region of the Orion Nebula. The cubes we provide
are among the largest 3D spectroscopic mosaics created so far.
We showed that the MUSE data are of high quality in terms of positional
accuracy, atmospheric refraction correction, $V$-band magnitude reconstruction,
velocities, and flux calibration. We also pointed out artifacts in and
imperfections of the data, and explained why two representations in cube form
are necessary to cover all possible investigations.

The ensemble of data presented here already allows to investigate a variety of
science topics.
In this paper, we restricted ourselved to only give a demonstration, with a
simple analysis of the ionized gas in the nebula. We derived the extinction
towards the ionization front, and the electron temperature and density using
two different emission line ratios, showing the physical properties in
different layers of the warm gas around $\theta^1$\,Ori\,C and the Trapezium
cluster.
Further analysis of this data will be presented in McLeod et al.\ (in prep.)
where we will analyze structures and kinematics in the Orion Nebula.

Note that smart spatial binning of the existing MUSE data would enable
detection weaker spectral features that in our pixel-by-pixel analysis are
detected with sufficient S/N only in the central part.  In the long term it
might prove useful to re-observe the field with MUSE, with possibly a longer
exposure time and/or more exposures per pointing, maybe mapping an even larger
field. This would allow to estimate physical properties from fainter emission
lines, and to map properties with even lower systematic effects to the
outskirts of M\,42.

Until that time, the current dataset is of high quality and already maps the
most interesting area of the Orion Nebula in the optical wavelength range.  It
can serve as reference for many follow-up studies.
To enable the widest possible use of these exceptional legacy data for
science questions from members of the community, we publicly release the
complete, reduced cubes as well as the inferred reddening, density, and
temperature maps on \url{http://muse-vlt.eu/science}.

\begin{acknowledgements}
The authors thank the (rest of) the MUSE consortium, the teams that conducted
first light and commissioning observations, and support from ESO during these
activities.
We also thank C.~R.\ O'Dell and G.\ Ferland for helpful comments.
We thank our referee, W.~Henney for an insightful and detailed report, that
enabled us to improve several critical points.
PMW and SK received funding through BMBF Verbundforschung (project MUSE-AO,
grant 05A14BAC and 05A14MGA).
AMI acknowledges support from Agence Nationale de la Recherche through the
STILISM project (ANR-12-BS05-0016-02).
RB acknowledges support from the ERC advanced grant 339659-MUSICOS.
We are grateful to the developers of software such as SAOImage DS9 (developed
by Smithsonian Astrophysical Observatory, \url{http://ds9.si.edu/}), IRAF,
PyNeb, APLpy (\url{http://aplpy.github.com}), and topcat
(\url{http://www.starlink.ac.uk/topcat/}), without which work on this data
would have been much more difficult.
We also thank the AstroPy community for useful discussions.
This research has made use of the SIMBAD database, operated at CDS, Strasbourg,
France.
\end{acknowledgements}

\bibliographystyle{aa}
\bibliography{orion.bib}

\begin{thebibliography}{69}
\expandafter\ifx\csname natexlab\endcsname\relax\def\natexlab#1{#1}\fi

\bibitem[{{Alonso-Herrero} {et~al.}(2009){Alonso-Herrero},
  {Garc{\'{\i}}a-Mar{\'{\i}}n}, {Monreal-Ibero}, {Colina}, {Arribas},
  {Alfonso-Garz{\'o}n}, \& {Labiano}}]{AlonsoHerrero09}
{Alonso-Herrero}, A., {Garc{\'{\i}}a-Mar{\'{\i}}n}, M., {Monreal-Ibero}, A.,
  {et~al.} 2009, A\&A, 506, 1541

\bibitem[{{Arthur} \& {Hoare}(2006)}]{2006ApJS..165..283A}
{Arthur}, S.~J. \& {Hoare}, M.~G. 2006, ApJS, 165, 283

\bibitem[{Bacon {et~al.}(2014)Bacon, Vernet, Borisova, Bouch\'e, Brinchmann,
  Carollo, Carton, Caruana, Cerda, Contini, Franx, Girard, Guerou, Haddad, Hau,
  Herenz, Herrera, Husemann, Husser, Jarno, Kamann, Krajnovic, Lilly, Mainieri,
  Martinsson, Palsa, Patricio, P\'econtal, Pello, Piqueras, Richard, Sandin,
  Schroetter, Selman, Shirazi, Smette, Soto, Streicher, Urrutia, Weilbacher,
  Wisotzki, \& Zins}]{2014Msngr.157...13B}
Bacon, R., Vernet, J., Borisova, E., {et~al.} 2014, The Messenger, 157, 13

\bibitem[{{Baldwin} {et~al.}(1991){Baldwin}, {Ferland}, {Martin}, {Corbin},
  {Cota}, {Peterson}, \& {Slettebak}}]{BFM+91}
{Baldwin}, J.~A., {Ferland}, G.~J., {Martin}, P.~G., {et~al.} 1991, ApJ, 374,
  580

\bibitem[{{Baldwin} {et~al.}(2000){Baldwin}, {Verner}, {Verner}, {Ferland},
  {Martin}, {Korista}, \& {Rubin}}]{BVV+00}
{Baldwin}, J.~A., {Verner}, E.~M., {Verner}, D.~A., {et~al.} 2000, ApJS, 129,
  229

\bibitem[{{Bally} {et~al.}(2015){Bally}, {Ginsburg}, {Silvia}, \&
  {Youngblood}}]{2015arXiv150204711B}
{Bally}, J., {Ginsburg}, A., {Silvia}, D., \& {Youngblood}, A. 2015, A\&A

\bibitem[{{Bally} {et~al.}(2000){Bally}, {O'Dell}, \&
  {McCaughrean}}]{2000AJ....119.2919B}
{Bally}, J., {O'Dell}, C.~R., \& {McCaughrean}, M.~J. 2000, AJ, 119, 2919

\bibitem[{{Blagrave} {et~al.}(2007){Blagrave}, {Martin}, {Rubin}, {Dufour},
  {Baldwin}, {Hester}, \& {Walter}}]{2007ApJ...655..299B}
{Blagrave}, K.~P.~M., {Martin}, P.~G., {Rubin}, R.~H., {et~al.} 2007, ApJ, 655,
  299

\bibitem[{{Bohlin} {et~al.}(1978){Bohlin}, {Savage}, \&
  {Drake}}]{1978ApJ...224..132B}
{Bohlin}, R.~C., {Savage}, B.~D., \& {Drake}, J.~F. 1978, ApJ, 224, 132

\bibitem[{Cair\'os {et~al.}(2015)Cair\'os, Caon, \& Weilbacher}]{CCW15}
Cair\'os, L., Caon, N., \& Weilbacher, P. 2015, A\&A, in press

\bibitem[{{Cardelli} {et~al.}(1989){Cardelli}, {Clayton}, \& {Mathis}}]{CCM89}
{Cardelli}, J.~A., {Clayton}, G.~C., \& {Mathis}, J.~S. 1989, ApJ, 345, 245

\bibitem[{{Clegg} {et~al.}(1999){Clegg}, {Miller}, {Storey}, \&
  {Kisielius}}]{1999A&AS..135..359C}
{Clegg}, R.~E.~S., {Miller}, S., {Storey}, P.~J., \& {Kisielius}, R. 1999,
  A\&AS, 135, 359

\bibitem[{{Da Rio} {et~al.}(2009){Da Rio}, {Robberto}, {Soderblom}, {Panagia},
  {Hillenbrand}, {Palla}, \& {Stassun}}]{2009ApJS..183..261D}
{Da Rio}, N., {Robberto}, M., {Soderblom}, D.~R., {et~al.} 2009, ApJS, 183, 261

\bibitem[{{De Robertis} {et~al.}(1987){De Robertis}, {Dufour}, \&
  {Hunt}}]{1987JRASC..81..195D}
{De Robertis}, M.~M., {Dufour}, R.~J., \& {Hunt}, R.~W. 1987, JRASC, 81, 195

\bibitem[{{Doi} {et~al.}(2010){Doi}, {Tanaka}, {Fukugita}, {Gunn}, {Yasuda},
  {Ivezi{\'c}}, {Brinkmann}, {de Haars}, {Kleinman}, {Krzesinski}, \& {French
  Leger}}]{2010AJ....139.1628D}
{Doi}, M., {Tanaka}, M., {Fukugita}, M., {et~al.} 2010, AJ, 139, 1628

\bibitem[{{Doi} {et~al.}(2004){Doi}, {O'Dell}, \&
  {Hartigan}}]{2004AJ....127.3456D}
{Doi}, T., {O'Dell}, C.~R., \& {Hartigan}, P. 2004, AJ, 127, 3456

\bibitem[{{Ducati}(2002)}]{2002yCat.2237....0D}
{Ducati}, J.~R. 2002, VizieR Online Data Catalog, 2237, 0

\bibitem[{{Filippenko}(1982)}]{1982PASP...94..715F}
{Filippenko}, A.~V. 1982, PASP, 94, 715

\bibitem[{{Foukal}(1974)}]{1974PASP...86..211F}
{Foukal}, P. 1974, PASP, 86, 211

\bibitem[{{Garc{\'{\i}}a-D{\'{\i}}az} \& {Henney}(2007)}]{2007AJ....133..952G}
{Garc{\'{\i}}a-D{\'{\i}}az}, M.~T. \& {Henney}, W.~J. 2007, AJ, 133, 952

\bibitem[{{Garc{\'{\i}}a-D{\'{\i}}az}
  {et~al.}(2008){Garc{\'{\i}}a-D{\'{\i}}az}, {Henney}, {L{\'o}pez}, \&
  {Doi}}]{2008RMxAA..44..181G}
{Garc{\'{\i}}a-D{\'{\i}}az}, M.~T., {Henney}, W.~J., {L{\'o}pez}, J.~A., \&
  {Doi}, T. 2008, \rmxaa, 44, 181

\bibitem[{{Garc{\'{\i}}a-Mar{\'{\i}}n}
  {et~al.}(2009){Garc{\'{\i}}a-Mar{\'{\i}}n}, {Colina}, {Arribas}, \&
  {Monreal-Ibero}}]{GarciaMarin09}
{Garc{\'{\i}}a-Mar{\'{\i}}n}, M., {Colina}, L., {Arribas}, S., \&
  {Monreal-Ibero}, A. 2009, A\&A, 505, 1319

\bibitem[{{Graham} {et~al.}(2003){Graham}, {Meaburn}, \&
  {Redman}}]{2003MNRAS.343..419G}
{Graham}, M.~F., {Meaburn}, J., \& {Redman}, M.~P. 2003, MNRAS, 343, 419

\bibitem[{{Hillenbrand}(1997)}]{1997AJ....113.1733H}
{Hillenbrand}, L.~A. 1997, AJ, 113, 1733

\bibitem[{{Israel}(1978)}]{1978A&A....70..769I}
{Israel}, F.~P. 1978, A\&A, 70, 769

\bibitem[{{Johnstone} \& {Bally}(1999)}]{1999ApJ...510L..49J}
{Johnstone}, D. \& {Bally}, J. 1999, ApJL, 510, L49

\bibitem[{{Johnstone} \& {Bally}(2006)}]{2006ApJ...653..383J}
{Johnstone}, D. \& {Bally}, J. 2006, ApJ, 653, 383

\bibitem[{{Kamann} {et~al.}(2013){Kamann}, {Wisotzki}, \& {Roth}}]{KWR13}
{Kamann}, S., {Wisotzki}, L., \& {Roth}, M.~M. 2013, A\&A, 549, A71

\bibitem[{{Kelz} {et~al.}(2006){Kelz}, {Verheijen}, {Roth}, {Bauer}, {Becker},
  {Paschke}, {Popow}, {S{\'a}nchez}, \& {Laux}}]{2006PASP..118..129K}
{Kelz}, A., {Verheijen}, M.~A.~W., {Roth}, M.~M., {et~al.} 2006, PASP, 118, 129

\bibitem[{Kramida {et~al.}(2014)Kramida, {Yu.~Ralchenko}, Reader, \& {and NIST
  ASD Team}}]{NIST_ASD_2014}
Kramida, A., {Yu.~Ralchenko}, Reader, J., \& {and NIST ASD Team}. 2014, {NIST
  Atomic Spectra Database (v5.2), [\tt{http://physics.nist.gov/asd}]}

\bibitem[{{Luridiana} {et~al.}(2013){Luridiana}, {Morisset}, \&
  {Shaw}}]{2013ascl.soft04021L}
{Luridiana}, V., {Morisset}, C., \& {Shaw}, R.~A. 2013, {PyNeb: Analysis of
  emission lines}, Astrophysics Source Code Library, record ascl:1304.021

\bibitem[{{Luridiana} {et~al.}(2015){Luridiana}, {Morisset}, \& {Shaw}}]{PyNeb}
{Luridiana}, V., {Morisset}, C., \& {Shaw}, R.~A. 2015, A\&A, 573, A42

\bibitem[{{Mesa-Delgado} {et~al.}(2008){Mesa-Delgado}, {Esteban}, \&
  {Garc{\'{\i}}a-Rojas}}]{2008ApJ...675..389M}
{Mesa-Delgado}, A., {Esteban}, C., \& {Garc{\'{\i}}a-Rojas}, J. 2008, ApJ, 675,
  389

\bibitem[{Mesa-Delgado {et~al.}(2012)Mesa-Delgado, N\'u\~nez D\'{\i}az,
  Esteban, Garc\'{\i}a-Rojas, Flores-Fajardo, L\'opez-Mart\'{\i}n, Tsamis, \&
  Henney}]{MesaDelgado12}
Mesa-Delgado, A., N\'u\~nez D\'{\i}az, M., Esteban, C., {et~al.} 2012, MNRAS,
  426, 614

\bibitem[{{Mesa-Delgado} {et~al.}(2011){Mesa-Delgado},
  {N{\'u}{\~n}ez-D{\'{\i}}az}, {Esteban}, {L{\'o}pez-Mart{\'{\i}}n}, \&
  {Garc{\'{\i}}a-Rojas}}]{MesaDelgado11}
{Mesa-Delgado}, A., {N{\'u}{\~n}ez-D{\'{\i}}az}, M., {Esteban}, C.,
  {L{\'o}pez-Mart{\'{\i}}n}, L., \& {Garc{\'{\i}}a-Rojas}, J. 2011, MNRAS, 417,
  420

\bibitem[{{Noll} {et~al.}(2014{\natexlab{a}}){Noll}, {Kausch}, {Kimeswenger},
  {Barden}, {Jones}, {Modigliani}, {Szyszka}, \& {Taylor}}]{skycorr}
{Noll}, S., {Kausch}, W., {Kimeswenger}, S., {et~al.} 2014{\natexlab{a}}, A\&A,
  567, A25

\bibitem[{{Noll} {et~al.}(2014{\natexlab{b}}){Noll}, {Kausch}, {Kimeswenger},
  {Barden}, {Jones}, {Modigliani}, {Szyszka}, \&
  {Taylor}}]{2014ascl.soft08007N}
{Noll}, S., {Kausch}, W., {Kimeswenger}, S., {et~al.} 2014{\natexlab{b}},
  {Skycorr: Sky emission subtraction for observations without plain sky
  information}, Astrophysics Source Code Library

\bibitem[{{N{\'u}{\~n}ez-D{\'{\i}}az}
  {et~al.}(2013){N{\'u}{\~n}ez-D{\'{\i}}az}, {Esteban}, \&
  {Mesa-Delgado}}]{2013hsa7.conf..594N}
{N{\'u}{\~n}ez-D{\'{\i}}az}, M., {Esteban}, C., \& {Mesa-Delgado}, A. 2013, in
  Highlights of Spanish Astrophysics VII, ed. J.~C. {Guirado}, L.~M. {Lara},
  V.~{Quilis}, \& J.~{Gorgas}, 594--599

\bibitem[{{N{\'u}{\~n}ez-D{\'{\i}}az}
  {et~al.}(2012){N{\'u}{\~n}ez-D{\'{\i}}az}, {Mesa-Delgado}, {Esteban},
  {L{\'o}pez-Mart{\'{\i}}n}, {Garc{\'{\i}}a-Rojas}, \&
  {Luridiana}}]{2012MNRAS.421.3399N}
{N{\'u}{\~n}ez-D{\'{\i}}az}, M., {Mesa-Delgado}, A., {Esteban}, C., {et~al.}
  2012, MNRAS, 421, 3399

\bibitem[{{O'Dell}(2001)}]{2001ARA&A..39...99O}
{O'Dell}, C.~R. 2001, \araa, 39, 99

\bibitem[{{O'Dell} \& {Harris}(2010)}]{ODH10}
{O'Dell}, C.~R. \& {Harris}, J.~A. 2010, AJ, 140, 985

\bibitem[{{O'Dell} \& {Henney}(2008)}]{2008AJ....136.1566O}
{O'Dell}, C.~R. \& {Henney}, W.~J. 2008, AJ, 136, 1566

\bibitem[{{O'Dell} {et~al.}(2009){O'Dell}, {Henney}, {Abel}, {Ferland}, \&
  {Arthur}}]{2009AJ....137..367O}
{O'Dell}, C.~R., {Henney}, W.~J., {Abel}, N.~P., {Ferland}, G.~J., \& {Arthur},
  S.~J. 2009, AJ, 137, 367

\bibitem[{{O'Dell} \& {Wen}(1994)}]{1994ApJ...436..194O}
{O'Dell}, C.~R. \& {Wen}, Z. 1994, ApJ, 436, 194

\bibitem[{{O'Dell} \& {Yusef-Zadeh}(2000)}]{2000AJ....120..382O}
{O'Dell}, C.~R. \& {Yusef-Zadeh}, F. 2000, AJ, 120, 382

\bibitem[{{Osterbrock} \& {Ferland}(2005)}]{OF05}
{Osterbrock}, D.~E. \& {Ferland}, G.~J. 2005, {Astrophysics of Gaseous Nebulae
  and Active Galactic Nuclei, Second Edition} ({University Science Books})

\bibitem[{{Osterbrock} {et~al.}(1992){Osterbrock}, {Tran}, \&
  {Veilleux}}]{1992ApJ...389..305O}
{Osterbrock}, D.~E., {Tran}, H.~D., \& {Veilleux}, S. 1992, ApJ, 389, 305

\bibitem[{{Patat} {et~al.}(2011){Patat}, {Moehler}, {O'Brien}, {Pompei},
  {Bensby}, {Carraro}, {de Ugarte Postigo}, {Fox}, {Gavignaud}, {James},
  {Korhonen}, {Ledoux}, {Randall}, {Sana}, {Smoker}, {Stefl}, \&
  {Szeifert}}]{2011A&A...527A..91P}
{Patat}, F., {Moehler}, S., {O'Brien}, K., {et~al.} 2011, A\&A, 527, A91

\bibitem[{{Pogge} {et~al.}(1992){Pogge}, {Owen}, \&
  {Atwood}}]{1992ApJ...399..147P}
{Pogge}, R.~W., {Owen}, J.~M., \& {Atwood}, B. 1992, ApJ, 399, 147

\bibitem[{{Robberto} {et~al.}(2013){Robberto}, {Soderblom}, {Bergeron},
  {Kozhurina-Platais}, {Makidon}, {McCullough}, {McMaster}, {Panagia}, {Reid},
  {Levay}, {Frattare}, {Da Rio}, {Andersen}, {O'Dell}, {Stassun}, {Simon},
  {Feigelson}, {Stauffer}, {Meyer}, {Reggiani}, {Krist}, {Manara},
  {Romaniello}, {Hillenbrand}, {Ricci}, {Palla}, {Najita}, {Ananna},
  {Scandariato}, \& {Smith}}]{2013ApJS..207...10R}
{Robberto}, M., {Soderblom}, D.~R., {Bergeron}, E., {et~al.} 2013, ApJS, 207,
  10

\bibitem[{{Rosado} {et~al.}(2001){Rosado}, {de la Fuente}, {Arias}, {Raga}, \&
  {Le Coarer}}]{2001AJ....122.1928R}
{Rosado}, M., {de la Fuente}, E., {Arias}, L., {Raga}, A., \& {Le Coarer}, E.
  2001, AJ, 122, 1928

\bibitem[{{Samus} {et~al.}(2009){Samus}, {Durlevich}, \& {et
  al.}}]{2009yCat....102025S}
{Samus}, N.~N., {Durlevich}, O.~V., \& {et al.} 2009, VizieR Online Data
  Catalog, 1, 2025

\bibitem[{{S{\'a}nchez} {et~al.}(2007){S{\'a}nchez}, {Cardiel}, {Verheijen},
  {Mart{\'{\i}}n-Gord{\'o}n}, {Vilchez}, \& {Alves}}]{SCV+07}
{S{\'a}nchez}, S.~F., {Cardiel}, N., {Verheijen}, M.~A.~W., {et~al.} 2007,
  A\&A, 465, 207

\bibitem[{Shields(1990)}]{Shi90}
Shields, G. 1990, ARA\&A, 28, 525

\bibitem[{{Skrutskie} {et~al.}(2006){Skrutskie}, {Cutri}, {Stiening},
  {Weinberg}, {Schneider}, {Carpenter}, {Beichman}, {Capps}, {Chester},
  {Elias}, {Huchra}, {Liebert}, {Lonsdale}, {Monet}, {Price}, {Seitzer},
  {Jarrett}, {Kirkpatrick}, {Gizis}, {Howard}, {Evans}, {Fowler}, {Fullmer},
  {Hurt}, {Light}, {Kopan}, {Marsh}, {McCallon}, {Tam}, {Van Dyk}, \&
  {Wheelock}}]{2MASS}
{Skrutskie}, M.~F., {Cutri}, R.~M., {Stiening}, R., {et~al.} 2006, AJ, 131,
  1163

\bibitem[{{Smette} {et~al.}(2015{\natexlab{a}}){Smette}, {Kausch}, {Sana},
  {Noll}, {Horst}, {Kimeswenger}, {Barden}, {Szyszka}, {Jones}, {Gallene},
  {Vinther}, {Ballester}, \& {Kerber}}]{2015ascl.soft01013S}
{Smette}, A., {Kausch}, W., {Sana}, H., {et~al.} 2015{\natexlab{a}}, {Molecfit:
  Telluric absorption correction tool}, Astrophysics Source Code Library

\bibitem[{{Smette} {et~al.}(2015{\natexlab{b}}){Smette}, {Sana}, {Noll},
  {Horst}, {Kausch}, {Kimeswenger}, {Barden}, {Szyszka}, {Jones}, {Gallenne},
  {Vinther}, {Ballester}, \& {Taylor}}]{molecfit}
{Smette}, A., {Sana}, H., {Noll}, S., {et~al.} 2015{\natexlab{b}}, A\&A, 576,
  A77

\bibitem[{{Storey} \& {Hummer}(1995)}]{1995MNRAS.272...41S}
{Storey}, P.~J. \& {Hummer}, D.~G. 1995, MNRAS, 272, 41

\bibitem[{{Str{\"o}mgren}(1939)}]{1939ApJ....89..526S}
{Str{\"o}mgren}, B. 1939, ApJ, 89, 526

\bibitem[{{Takahashi} {et~al.}(2013){Takahashi}, {Ho}, {Teixeira}, {Zapata}, \&
  {Su}}]{2013ApJ...763...57T}
{Takahashi}, S., {Ho}, P.~T.~P., {Teixeira}, P.~S., {Zapata}, L.~A., \& {Su},
  Y.-N. 2013, ApJ, 763, 57

\bibitem[{{Takami} {et~al.}(2002){Takami}, {Usuda}, {Sugai}, {Suto}, {Pyo},
  {Takeyama}, {Aoki}, {Mizutani}, \& {Tanaka}}]{2002ApJ...566..910T}
{Takami}, M., {Usuda}, T., {Sugai}, H., {et~al.} 2002, ApJ, 566, 910

\bibitem[{Tsamis {et~al.}(2013)Tsamis, Flores-Fajardo, Henney, Walsh, \&
  Mesa-Delgado}]{Tsamis13}
Tsamis, Y.~G., Flores-Fajardo, N., Henney, W.~J., Walsh, J.~R., \&
  Mesa-Delgado, A. 2013, MNRAS, 430, 3406

\bibitem[{{Tsamis} \& {Walsh}(2011)}]{Tsamis11}
{Tsamis}, Y.~G. \& {Walsh}, J.~R. 2011, MNRAS, 417, 2072

\bibitem[{{van der Werf} {et~al.}(2013){van der Werf}, {Goss}, \&
  {O'Dell}}]{2013ApJ...762..101V}
{van der Werf}, P.~P., {Goss}, W.~M., \& {O'Dell}, C.~R. 2013, ApJ, 762, 101

\bibitem[{{Vasconcelos} {et~al.}(2005){Vasconcelos}, {Cerqueira}, {Plana},
  {Raga}, \& {Morisset}}]{Vasconcelos05}
{Vasconcelos}, M.~J., {Cerqueira}, A.~H., {Plana}, H., {Raga}, A.~C., \&
  {Morisset}, C. 2005, AJ, 130, 1707

\bibitem[{Weilbacher {et~al.}(2003)Weilbacher, Duc, \& Fritze-von
  Alvensleben}]{WDF03}
Weilbacher, P.~M., Duc, P.-A., \& Fritze-von Alvensleben, U. 2003, A\&A, 397,
  545

\bibitem[{{Weilbacher} {et~al.}(2012){Weilbacher}, {Streicher}, {Urrutia},
  {Jarno}, {P{\'e}contal-Rousset}, {Bacon}, \& {B{\"o}hm}}]{WSU+12}
{Weilbacher}, P.~M., {Streicher}, O., {Urrutia}, T., {et~al.} 2012, in
  Proc.~{SPIE}, Vol. 8451, {Software and Cyberinfrastructure for Astronomy II}

\bibitem[{{Yusef-Zadeh}(1990)}]{1990ApJ...361L..19Y}
{Yusef-Zadeh}, F. 1990, ApJL, 361, L19

\bibitem[{{Zuckerman}(1973)}]{1973ApJ...183..863Z}
{Zuckerman}, B. 1973, ApJ, 183, 863

\end{thebibliography}

\end{document}